\documentclass[12pt,reqno]{amsart}
\usepackage{times}
\usepackage{amsmath, amssymb, graphicx, url, algorithm2e,mathtools}
\usepackage[round]{natbib}
\usepackage{xcolor}
\usepackage{wrapfig}
\usepackage{tikz}
\tikzset{every node/.style={scale=0.8}}
\usepackage{multirow}
\usepackage{mathptmx}
\bibstyle{natbib}
\allowdisplaybreaks
\usepackage[left=1in,right=1in,top=1in,bottom=1in]{geometry}

\usepackage{color-edits}
\addauthor{vs}{red}

\usepackage[utf8]{inputenc} 
\usepackage[T1]{fontenc}    
\usepackage[hidelinks]{hyperref} 
\usepackage{url}            
\usepackage{booktabs}       
\usepackage{amsfonts}       
\usepackage{nicefrac}       
\usepackage{microtype}      

\usepackage{caption}
\usepackage{subcaption}
\usepackage{ragged2e}

\newcommand{\Ep}{{\mathrm{E}}}

\renewcommand{\Pr}{{\mathrm{P}}}

\newcommand{\divg}{\mathrm{div}}

\newcommand{\indep}{\perp\!\!\!\!\perp}

\DeclareMathOperator{\Var}{Var}
\DeclareMathOperator{\Odds}{Odds}

\newtheorem{Theorem}{Theorem}

\newtheorem{Lemma}{Lemma}

\newtheorem{assumption}{Assumption}
\newtheorem{definition}{Definition}

\theoremstyle{definition}
\newtheorem{remark}{Remark}
\newtheorem{example}{Example}

\title[]{Long Story Short:\\ Omitted Variable Bias in Causal Machine Learning}
\dedicatory{\normalfont\small Forthcoming in the Review of Economics and Statistics.\\
\url{https://doi.org/10.1162/REST.a.1705}}

\author[V. Chernozhukov]{Victor Chernozhukov\textsuperscript{\dag}}
\thanks{\textsuperscript{\dag} Dept. of Economics, Massachusetts Institute of Technology, Cambridge, MA, USA. Email: \href{mailto:vchern@mit.edu}{vchern@mit.edu}}
\author[C. Cinelli]{Carlos Cinelli\textsuperscript{*}}  
\thanks{\textsuperscript{*} Dept. of Statistics, University of Washington, Seattle, WA, USA. Email: \href{mailto:cinelli@uw.edu}{cinelli@uw.edu}}
\author[W. Newey]{Whitney K. Newey\textsuperscript{\ddag}}  
\thanks{\textsuperscript{\ddag} Dept. of Economics, Massachusetts Institute of Technology, Cambridge, MA, USA. Email: \href{mailto:wnewey@mit.edu}{wnewey@mit.edu}}
\author[A. Sharma]{Amit Sharma\textsuperscript{$\|$}} 
\thanks{\textsuperscript{$\|$} Microsoft Research India, Bangalore, India. Email: \href{mailto:amshar@microsoft.com}{amshar@microsoft.com}}
\author[V. Syrgkanis]{Vasilis Syrgkanis\textsuperscript{\S}}
\thanks{\textsuperscript{\S} Dept. of Mgmt Science and Engineering, Stanford University, Stanford, CA, USA. Email: \href{mailto:vsyrgk@stanford.edu}{vsyrgk@stanford.edu}}

\begin{document}

\begin{abstract}
We develop a general theory of omitted variable bias for a wide range of common causal parameters, including average treatment effects, average causal derivatives, and policy effects from covariate shifts. We show how plausibility judgments on the maximum explanatory power of omitted variables are sufficient to bound the bias, facilitating sensitivity analysis in otherwise complex models. Finally, we provide statistical inference methods that can leverage modern machine learning algorithms for estimation.  These results allow empirical researchers to perform sensitivity analyses in a flexible class of machine-learned causal models using very simple tools. Empirical examples demonstrate the utility of our approach.
\end{abstract}					

\maketitle

\section{Introduction}

Unmeasured confounding is a pervasive issue in studies that aim to draw causal inferences from observational data. Such studies typically rely on a conditional ignorability (also known as unconfoundedness) assumption, which states that the treatment assignment is independent of potential outcomes given a set of observed covariates \citep{Rosenbaum1983, pearl:causality,AngristBook, imbens2015causal}. This assumption, however, requires that there are no unobserved confounders influencing both the treatment and the outcome. When such variables are omitted from the analysis, empirical estimates may differ from the true causal effect of interest, giving rise to what is now commonly known as ``omitted variable bias.''

The omitted variable bias (OVB) problem is one of the most significant threats to the identification of causal effects.  
In the context of linear models, this bias amounts to the difference between the coefficients of the treatment variable from two distinct outcome regressions: one that controls only for observed covariates (the ``short'' regression) and another that would additionally control for unobserved variables (the ``long'' regression).
Formulas characterizing this difference play a foundational role 
in statistics, econometrics, and related fields (see, e.g., discussions in classical and modern textbooks, such as  \citealp{goldberger1991course, AngristBook} and \citealp{wooldridge:text}).

But while linear models are widely used in applied work, they are often overly restrictive. For example, in the binary treatment case, using linear models when treatment effects are heterogeneous may yield unintuitive or even misleading estimates of the causal effects of interest \citep{aronow2016does,sloczynski2022interpreting}. 
To address these limitations, many empirical analysts have turned their attention to more flexible nonlinear or nonparametric models, often leveraging modern machine learning techniques for estimation and inference \citep{vanderlaan:book,BCFH2013,DML,athey:grf}.
These tools offer the flexibility to capture complex relationships between variables, avoiding stringent functional form assumptions in causal effect estimation. 
Yet, though sensitivity analyses for nonparametric models do exist (see related literature below), we currently lack general results characterizing the form of omitted variable bias in nonlinear models, as we have for the linear case.  Our work aims to address this gap.

In this paper we develop a general theory of omitted variable bias for a wide range of common causal parameters  that can be identified as linear functionals of the conditional expectation function (CEF) of the outcome. Such functionals encompass many (if not most) of the traditional targets of investigation in causal inference studies, such as averages of potential outcomes, average treatment effects, average causal derivatives, and policy effects from covariate shifts. We  allow for arbitrary treatment (e.g., continuous or binary) and outcome variables. Our theory applies to general nonparametric models, while naturally allowing for \mbox{(semi-)parametric} restrictions (such as partial linearity) when such assumptions are made.  Our formulation recovers well-known and familiar OVB results for linear models as a special case, and it can be seen as its natural generalization to nonlinear models. Importantly, we show that the general nonparametric formula of the bias still has a simple and interpretable form.

More specifically, we first formalize the OVB problem in the nonparametric setting. Paralleling the linear case, we define the OVB as the difference between the ``short'' and ``long'' functionals of the outcome regression, where the former omits and the latter includes the latent variables. To derive the OVB, our construction then leverages the Riesz-Frechet representation of the target functionals, which allows us to rewrite the parameters of interest as weighted averages of the outcome regression, with weights given by the Riesz representers (RRs).\footnote{The RRs are also important in asymptotic variance formulae for estimators of parameters of interest \citep{newey1994}.} We show that the OVB arises as a by-product of confounders introducing systematic errors in both the outcome regression and in the RRs for the parameter of interest. Furthermore, plausibility judgments on the maximum explanatory power of latent variables suffice to place overall bounds on the bias, simplifying the task of sensitivity analysis even when using nonparametric or otherwise complex models.

 Although these results may initially seem abstract to those not familiar with Riesz representation theory, in many leading examples the RRs in fact correspond to quantities that are well-known to empirical researchers. For instance, when estimating the average treatment effect in a partially linear model, the RR is the (variance scaled) residualized treatment, after ``partialling out'' the control covariates. Or, when estimating an average treatment effect in a general nonparametric model with a binary treatment, the RR is now given by another familiar quantity---the inverse probability of treatment weights (IPTW). In such cases, we show that the bias can be reparameterized in terms of simple percentage gains in variance explained (or precision) in the treatment and the outcome regression due to unmeasured confounders, again facilitating the interpretation and use of the OVB formulas in practice.  We further help analysts make plausibility judgments on the magnitude of sensitivity parameters by means of comparison of the relative strength of unobserved confounders against the strength of observed covariates.

Finally, we provide statistical inference using debiased machine learning (DML) and auto-DML  \citep{DML, RR-local, Adverse, AutoDML-Lasso}. Our construction makes it possible to use modern machine learning methods for estimating the identifiable components of the omitted variable bias formula, including regression functions, Riesz representers, the norm of regression residuals, and the norm of Riesz representers. These results enable flexible and efficient statistical inference on the bounds of the target parameter, allowing researchers to perform sensitivity analyses against unmeasured confounding in a flexible class of machine-learned causal models using simple and interpretable tools.

\subsection*{Related Literature}

Our work is most closely related to the literature that derives OVB formulas for linear models, such as those found in traditional textbooks  and recent extensions \citep{goldberger1991course,AngristBook,wooldridge:text,frank:smr2000,oster2019unobservable,CH2020,cinelli2020omitted}.  We advance this literature by providing analogous, easily explainable OVB formulas for a broad and rich class of causal parameters, all for general nonlinear models, \emph{with or without further parametric restrictions}. Importantly, we provide a single unifying framework that covers all these cases, and that can be easily specialized depending on the target parameter and on whether additional parametric assumptions (if any) are made. We further advance the OVB literature by providing flexible and efficient statistical inference methods, leveraging modern machine learning algorithms with debiased machine learning.

More broadly, our work is related to the extensive literature on sensitivity analysis against unmeasured confounders. Here we highlight the key differences between our approach and existing methods, while relegating a more detailed review to the Online Appendix. First, many prior works on sensitivity analysis either focus exclusively on binary treatments (e.g., \citealp{rosenbaum:b1987,masten:e2018,zhao:jrssb2019,bonvini2021sensitivity}), target a single estimand of interest, such as a causal risk ratio \citep{ding:ecm2016,evalue}, or impose parametric assumptions on the observed data or on the nature of unobserved confounding \citep{rosenbaum1983assessing,imbens2003sensitivity,dorie2016flexible,cinelli:icml2019,zhang2021exploiting}. Our approach differs from these in that (i) it is not limited to binary treatments, (ii) it covers a broader range of target parameters, such as average causal derivatives and average policy effects from covariate shifts, and (iii) it does not require parametric assumptions on the observed data nor on the nature of confounding.

Even if we focus solely on the important special case of estimating an average treatment effect (ATE) or average treatment effect on the treated (ATT) with a binary treatment, our OVB results usefully complement other seminal approaches to this problem such as those of \cite{rosenbaum:b1987} or the marginal sensitivity models of \cite{tan:jasa2006}. Whereas such approaches limit the strength of confounding through its impact on the \emph{worst-case} change in the odds ratio of treatment assignment, our OVB approach shows that this assumption is much stronger than what is needed to bound the bias---all that is required by the OVB formula are claims about changes in \emph{averages} (see Section~\ref{sec:msm}). Moreover, even in stylized models of treatment assignment (e.g., a logistic model with a Gaussian latent confounder), worst-case approaches such as the ones in \cite{rosenbaum:b1987} and \cite{tan:jasa2006} have a naturally  unbounded sensitivity parameter, no matter how small the actual degree of confounding is (see the Online Appendix for an example).

Importantly, our OVB-based approach generally differs from traditional sensitivity analyses in that it derives the exact OVB formula for the target parameters we cover. For example, our results show that the bias of the ATE in the binary treatment case is \emph{not} determined by deviations on the odds of treatment; rather, it is determined by three quantities: (i) the maximum explanatory power of confounders in the treatment regression, as given by gains in precision, (ii) the maximum explanatory power of confounders in the outcome regression, as given by gains in variance explained, and (iii) by the correlation of errors in the regression function and the IPTW. Therefore, beyond being a tool for sensitivity analysis, OVB results such as ours provide a precise characterization of how different estimands respond to unmeasured confounding.

\subsection*{Overview of the paper} Section~\ref{sec:plm} presents our method in the simpler context of partially linear models. Section~\ref{sec:general} derives the main result of the paper, i.e., formalizes the omitted variable bias problem and derives its formula for the general class of continuous linear functionals of (projections of) the conditional expectation function of the outcome---all for general, nonparametric causal models. In Section~\ref{sec:inference} we construct high-quality methods for statistical inference under omitted variable bias, by leveraging recent advances in debiased machine learning with Riesz representers. Section~\ref{sec:empirical-examples} demonstrates the use of our tools to assess the robustness of causal claims in a detailed empirical example that estimates the average treatment effect of 401(k) eligibility on net financial assets. Section~\ref{sec:conclusion} concludes with suggestions for possible extensions. The Online Appendix contains all proofs, a more extensive literature review, as well as an additional empirical example that illustrates sensitivity analyses for average causal derivatives with continuous treatments.

\subsection*{Notation} All random vectors are defined on the probability space with probability measure~$\Pr$. We consider a random vector $Z=(Y,W)$ with distribution $P$  taking values $z$ in its support $\mathcal{Z}$;  we use $P_{V}$ to denote the probability law of any subvector $V$ and $\mathcal{V}$ to denote its support.  We use $\| f \|_{P,q} = \| f(Z)\|_{P,q}$ to denote the $L^q(P)$ norm of a measurable function $f: \mathcal{Z} \to \mathbb{R}$ and also the $L^q(P)$ norm of random variable $f(Z)$.  For a differentiable map $x \mapsto g(x)$, from $\mathbb{R}^d$ to $\mathbb{R}^k$,  $\partial_{x'} g$ abbreviates the partial derivatives  $(\partial/\partial x') g(x)$, and $\partial_{x'} g(x_0)$ means $\partial_{x'} g (x) \mid_{x = x_0}$.  We use $x'$ to denote the transpose of a column vector $x$. We use the conventional notation $dL/dP$ to denote the Radon-Nikodym derivative of measure $L$ with respect to $P$. 

We use $R^2_{V\sim U}$ to denote the (uncentered) $R^2$ from the orthogonal linear projection of a scalar random variable $V$ on a random vector $U$. 
We follow \cite{doksum1995nonparametric} and define the \emph{nonparametric} $R^2$ as
$$ 
\eta^2_{V\sim U} := \frac{\Var(\Ep[V\mid U])}{\Var(V)}
=  1- \frac{\Ep[\Var(V\mid U)]}{\Var(V)} =  R^2_{V-\Ep[V]\sim \Ep[V\mid U]-\Ep[V]},
$$
which also equals the linear $R^2$ of the projection of $V-\Ep[V]$ on $\Ep[V \mid U]-\Ep[V]$. The nonparametric \emph{partial} $R^2$ of $U$ with another variable $V$ \emph{given} $X$ measures the additional gain in the explanatory power that $U$ provides, beyond what is already explained by $X$. This equals the relative decrease in the average residual variance, or the linear $R^2$ of the projection of the residual variation of $V$ onto the residual variation of $\Ep[V\mid X,U]$:
\begin{equation}
\label{eq:partialr2}
\eta^2_{V\sim U\mid X} := 
 \frac{\eta^2_{V\sim UX} - \eta^2_{V\sim X}}{1 - \eta^2_{V\sim X}} 
 = 1-\frac{\Ep[\Var(V\mid X, U)]}{\Ep[\Var(V\mid X)]} =  R^2_{V -\Ep[V\mid X]\sim \Ep[V\mid X,U] - \Ep[V\mid X]} .
\end{equation}
For a binary random variable $V$, we use $\Odds(V\mid U):= P(V=1\mid U)/(1-P(V=1\mid U))$ to denote the conditional odds of $V$ given  random vector $U$. These metrics will be useful for characterizing the strength of unmeasured confounders in the next sections.

\section{Warm-Up: Omitted Variable Bias in Partially Linear Models}\label{sec:plm}

To fix ideas, we begin our discussion in the context of partially linear models (PLM). These results not only provide the key intuitions and the building blocks for the general case of nonseparable, nonparametric models of Section~\ref{sec:general}, but they are also important in their own right, as these models are widely used in applied~work.

\subsection{Problem set-up}
\label{sec:setup-plm}

Suppose an empirical researcher is  interested in the regression coefficient $\theta$ from a partially linear regression model of the following form
\begin{equation}\label{eq:PLM1}
Y = \theta D + f(X, U) + \epsilon, 
\end{equation}
where here $Y$ denotes a real-valued outcome, $D$ a real-valued treatment, $X$ an observed {vector} of covariates, and $U$ an \emph{unobserved} vector of confounders. We refer to $W: = (D,X,U)$ as the ``long'' list of regressors, and to equation (\ref{eq:PLM1}) as the ``long'' regression.  This regression model need not be correctly specified, and the error term simply obeys the orthogonality condition $\Ep[ \epsilon (D- \Ep[D\mid X, U])]=0$.\footnote{That is $g(W):= \theta D + f(X, U)$ is the projection of the CEF on the space of functions that are partially linear in $D$.}

However, since $U$ is unobserved, the coefficient $\theta$ cannot be identified from the observed data. The researcher then naturally approximates $\theta$ by estimating the analogous regression coefficient $\theta_s$ from a partially linear model that \emph{omits} $U$:
\begin{equation}\label{eq:PLM2}
 Y = \theta_s D + f_s(X) + \epsilon_s.
 \end{equation}
As before, this model need not be correctly specified, and we simply have  $\Ep[ \epsilon_s (D- \Ep[D\mid X])]=0$. We call equation~(\ref{eq:PLM2}) the ``short regression,'' and $W_s:= (D,X)$ the ``short'' list of observed regressors. 

Evidently, in general  $\theta_s$ is not equal to $\theta$, and this naturally leads to the question of how far this ``proxy'' parameter deviates from the true inferential target. Our goal is to analyze the difference between the long and short parameters---the omitted variable bias (OVB):
$$
\theta - \theta_s,
$$
and perform inference on this bias under various hypotheses on the strength of the unobserved confounders~$U$.

\subsection{Omitted variable bias for partially linear models} 
\label{sec:ovb-plm}
Using a Frisch-Waugh-Lovell (FWL) partialling out argument (\citealp{frisch1933partial, lovell1963seasonal}),  one can  express the long and short parameters, $\theta$ and $\theta_s$, as the linear regression coefficients of $Y$ on the  residuals $D- \Ep[D\mid X, U]$ and $D- \Ep[D\mid X]$, respectively. 

That is,
\begin{equation}
\label{eq:partial_covariance}
\theta = \Ep Y \alpha (W),  \quad \quad \theta_s = \Ep Y \alpha_s(W_s);
\end{equation}
where we define
$$
\alpha(W) := \frac{D- \Ep[D\mid X, U]}{\Ep (D- \Ep[D\mid X, U ])^2 }, \quad \alpha_s(W_s) := \frac{D- \Ep[D\mid X]}{\Ep (D- \Ep[D\mid X])^2 }.
$$
For reasons that will become clear in the next section, we can refer to $\alpha(W)$ and $\alpha_s(W_s)$ as the ``long'' and ``short'' Riesz representers (RR). 

Now let $g(W) := \theta D + f(X, U)$ and $g_{s}(W_s) := \theta_s D + f_s(X)$ denote the long and short partially linear projections, respectively. Using the orthogonality conditions in (\ref{eq:PLM1}) and (\ref{eq:PLM2}), we can further express $\theta$ and $\theta_s$ as 
\begin{equation} 
\Ep Y \alpha (W) = \Ep g(W) \alpha(W),  \quad \quad  \Ep Y \alpha_s(W_s) = \Ep g_s(W_s) \alpha_s(W_s).
\end{equation}

\noindent Our first characterization of the OVB is thus as follows, where we use the shorthand notation: $g =g(W)$, $g_s = g_s(W_s)$, $\alpha=\alpha(W)$, and $\alpha_s = \alpha_s(W_s)$. 
\begin{Theorem}[\textbf{OVB---PLM}]
\label{thm:ovb-plm} 
Assume that  $Y$ and $D$ are square-integrable with \mbox{$\Ep(D- \Ep[D\mid X, U])^2>0.$}
Then the OVB for the partially linear regression coefficients of equation~(\ref{eq:partial_covariance}) is given by the covariance between the regression error and the RR error,
$$
\theta - \theta_s 
= \Ep (g -g_s) (\alpha - \alpha_s ).
$$
This bias  can be further reparameterized in terms of $R^2$ measures, 
$$
\theta - \theta_s = \underbracket{\mathrm{Cor} ( g-g_s, \alpha- \alpha_s)}_{\rho} \times \underbracket{\sqrt{R^2_{ Y- g_s \sim g- g_s}}}_{C_Y}\times \underbracket{\sqrt{ \frac{(1-R^2_{\alpha\sim \alpha_s})}{R^2_{\alpha\sim \alpha_s}}}}_{C_D}\times \underbracket{\sqrt{\Ep(Y-g_s)^2 \Ep\alpha_s^2}}_{S},
$$
where, 
$1-R^2_{\alpha \sim \alpha_s} = \eta^2_{D \sim U \mid X}$, and $E\alpha^2_s = \frac{1}{\Ep(D- \Ep[D\mid X])^2} = \frac{1}{\Ep[\Var(D\mid X)]}$. Moreover,
if both $\Ep[Y\mid D, X, U]$ and $\Ep [Y \mid  D,X]$ are partially linear in $D$---that is, if $g$ and $g_s$ are correctly specified---we also have $\Ep(Y-g_s)^2 = \Ep[\Var(Y\mid D, X)]$  and $R^2_{ Y- g_s \sim g- g_s} = \eta^2_{Y \sim U \mid DX}.$
\end{Theorem}  

This result for partially linear models is new and it naturally generalizes  traditional OVB formulas for linear models (see Online Appendix). Note the theorem first states the bias formula in terms of the Riesz representers $\alpha$ and $\alpha_s$, and then specializes its interpretations. This is intentional, and it aims to facilitate the transition to the general case of nonparametric models in the next section.

\subsection{Making sense of the OVB formula}
\label{sec:making-sense-plm}

Theorem~\ref{thm:ovb-plm} shows that the bias decomposes into the product of an identifiable scaling factor, $S$, and a non-identifiable bias factor, $\rho C_Y C_D$. The scaling factor $S = \sqrt{\frac{\Ep(Y-g_s)^2}{\Ep(D-\Ep[D\mid X])^2}}$ depends on the ratio of the observed residual variances of the treatment and the outcome. Intuitively, it shows that the more variation the observed covariates explain of the outcome, the less room there is for omitted variable bias to arise. On the other hand, the more variation the observed covariates explain of the treatment, the less residual variation of the treatment is being used to identify the target parameter---this, in its turn, \emph{inflates}  the potential biasing effect of omitted variables.

Moving to the bias factor  $\rho C_Y C_D$, this term decomposes the bias in terms of the capacity of confounders to induce errors in the outcome  and in the treatment equation, as well as on how systematically related these errors are. The sensitivity parameter $R^2_{ Y- g_s \sim g- g_s}$ in $C_Y$ captures the potential of the confounder to induce errors in the outcome regression, as measured by how much residual variation the confounder explains of the outcome. The sensitivity parameter $1-R^2_{\alpha\sim \alpha_s}=\eta^2_{D \sim U \mid X}$ in $C_D$ captures the potential of the confounder to induce errors in the treatment regression, as measured by the proportion of residual variation of the treatment explained by $U$. Collectively, the pair $(1-R^2_{\alpha\sim \alpha_s},~ R^2_{ Y- g_s \sim g- g_s})$ measures the \emph{maximum explanatory power} of the confounders, which sets an upper bound on how much bias $U$ \emph{could} produce. However, for $U$ to actually create bias, it is not sufficient for it to create errors in the treatment and the outcome equations---it must do so \emph{systematically}. This is captured by the correlation of errors $\rho$.\footnote{For an example, consider the model $D = U^2$, $Y = \theta D + U$, with $U\sim N(0,1)$. Here $\eta^2_{D\sim U\mid X}=\eta^2_{Y\sim U\mid D,X} = 1$, that is, the latent variable $U$ nonparametrically explains 100\% of the residual variation of the treatment and of the outcome. However, since $U^2$ is uncorrelated with $U$, we have $\rho = 0$, and omitting this confounder induces no bias in the estimation of $\theta$.}

\subsection{Using the OVB formula to bound the bias} 
\label{sec:bound-plm}
Given plausibility judgments on the maximum value that the components of the OVB formula can take, $|\rho| \leq \rho^{\max}$, $R^2_{ Y- g_s \sim g- g_s}\leq R^{2\max}_Y$ and $1-R^2_{\alpha\sim \alpha_s} \leq R^{2\max}_D$, we immediately obtain the  following  bound on the bias,
\begin{align}
\label{eq:bound}
|\theta - \theta_s| \leq \rho^{\max} \times \sqrt{R^{2\max}_{Y} \times \left(\frac{R^{2\max}_{D}}{1-R^{2\max}_{D}}\right)}\times S.    
\end{align}
Additionally, if investigators are unwilling to make any assumptions about $\rho^{\max}$ and $R^{2\max}_Y$ (i.e., setting them to the trivial bound of 1), then (\ref{eq:bound}) reduces to a simple, {single-parameter} bound\footnote{This is similar in spirit to \cite{manski1990nonparametric}, in which, absent further assumptions, we consider the worst case scenario.} 
$$
|\theta - \theta_s| \leq \sqrt{\frac{R^{2\max}_{D}}{1 - R^{2\max}_{D}}} \times S.
$$
Note how this simplifies the complexity of plausibility judgments for sensitivity analysis. Researchers need only reason about the maximum explanatory power that unobserved confounders have in explaining treatment variation in order to place bounds on the bias.  Should the researcher later choose to postulate the maximum explanatory power of confounders with the outcome via $R^{2\max}_{Y}$, or the degree to which the confounder is adversarial, via $\rho^{\max}$, the bounds become tighter.

Finally, one might wonder whether the bound (\ref{eq:bound}) is the tightest possible bound one could obtain given the model assumptions and the distribution of observed variables. Sometimes the answer is yes. For example, in the Online Appendix, we demonstrate that when the observed residuals are homoskedastic Gaussian, it is always possible to construct a latent confounder \(U\), compatible with the observed distribution, that achieves any combination of values \((|\rho|,~R^2_{Y-g_s \sim g-g_s},~ 1- R^2_{\alpha \sim \alpha_s})~\in~[0,1)^3\). This implies that, in such a scenario, the bound (\ref{eq:bound}) is always sharp and cannot be improved without additional assumptions. Other times, the observed distribution may impose restrictions on the values the triple can take, and the bound  (\ref{eq:bound}) may be tightened by leveraging such restrictions. See the Online Appendix for further discussion.

\section{Main Results: Omitted Variable Bias in Nonparametric Causal Models}\label{sec:general}

We now derive the main results of the paper, and characterize the size of the omitted variable bias for a broad class of causal parameters that can be identified as linear functionals of the conditional expectation function of the outcome. Although more abstract, the presentation of this section largely parallels the special case of partially linear models given in Section~\ref{sec:plm}.

\subsection{Problem set-up}
\label{sec:setup-general}

As a motivating example, consider any causal or structural equation model that implies the existence of potential outcomes, $Y(d)$, under the intervention that sets the treatment \emph{experimentally} to~$d$. Assume that \emph{consistency} holds, namely, observed outcomes equal to the potential outcome for the treatment actually received, $Y:=Y(D)$. Further, let the treatment assignment $D$ obey the ignorability (conditional exogeneity) condition:
\begin{equation} \label{eq: CE}
Y(d) \indep D \mid \{X, U\},
\end{equation}
which states that the realized treatment $D$ is independent of the potential outcomes, conditionally on observed covariates $X$ and \emph{unobserved} confounders $U$.  Under this set-up, and when $d$ is in the support of $D$ given $X$, $U$, we then have the following (well-known) identification result
$$
\Ep[Y (d) \mid X, U]= \Ep [Y(d) \mid D=d, X, U] = \Ep [Y \mid D=d, X, U] =: g(d,X,U),
$$
that is, the conditional average potential outcome coincides with the ``long'' regression function of $Y$ on $D$, $X$, and $U$. Therefore, we can identify various causal parameters---functionals of the average potential outcome---from the regression function. Important examples include: (i) the average treatment effect (ATE)
$$
\theta = \Ep[Y(1) - Y(0)] = \Ep [g(1,X,U) - g(0, X, U)],
$$
for the case of a binary treatment $D$; and, (ii) the average causal derivative (ACD)  
$$
\theta = \Ep\left [\partial_d \Ep [Y(D)\mid X, U] \right] = \Ep[\partial_d g(D,X,U)],
$$
for the case of a continuous treatment $D$.

In fact, our framework is considerably more general, and it covers any target parameter of the following form.

\begin{assumption}[\textbf{Target ``Long'' Parameter}] 
\label{asm:theta}
The target parameter $\theta$ is a continuous linear functional of a regression function $g$:
\begin{equation}
\theta :=  \Ep m(W, g), 
\end{equation}
where $g$ is the projection of the CEF $\Ep[Y\mid W]$ onto a closed linear subspace $\Gamma \subseteq L^2(P_W)$:
\begin{align}
    g :=~& \arg\min_{\gamma \in \Gamma} \Ep(Y-\gamma(W))^2 \equiv \arg\min_{\gamma \in \Gamma} \Ep (\Ep[Y\mid W] -\gamma(W))^2
\end{align}
and the mapping $\gamma \mapsto m(w; \gamma)$ is linear in $\gamma \in \Gamma$, and continuous in $\gamma\in \Gamma$ with respect to the $L^2(P_W)$~norm.  
\end{assumption}

\noindent This formulation covers the two previous examples with scores
$m(W, g) = g(1, X,U) - g(0, X,U)$ for the ATE and 
$m(W, g) = \partial_d g(D, X,U)$ for the ACD. Moreover, it covers settings where the true CEF does not lie in the subspace $\Gamma$, allowing for mis-specification. Hence, it enables omitted variable bias for coefficients in best (partially) linear models, without any assumption of (partial) linearity of the CEF, as in Section~\ref{sec:plm}.

The continuity condition holds under the regularity conditions provided in the remark below. 
\begin{remark}[\textbf{Regularity Conditions for ATE and ACD}] As regularity conditions for the ATE we assume $\Ep Y^2 < \infty$ and the weak overlap condition:
$$
\Ep [ P(D=1\mid X,U)^{-1} P(D=0\mid X,U)^{-1}]< \infty.
$$
As regularity conditions for the ACD we assume $\Ep Y^2< \infty$,  that the conditional density $d \mapsto f(d \mid  x,u)$ is  continuously differentiable on its support $\mathcal{D}_{x,u}$, the regression function $d \mapsto g(d,x,u)$ is continuously differentiable on $\mathcal{D}_{x,u}$, and we have that $f(d\mid x,u)$ vanishes whenever $d$ is on the boundary  of $\mathcal{D}_{x,u}$. The above needs to hold for all values $x$ and $u$ in the support of $(X,U)$. We also impose the bounded information assumption:
$$
\Ep (\partial_d \log f(D \mid X,U ))^2 < \infty. 
$$
We also assume that the expectations defining $\theta$ and $\theta_s$ are finite.
These conditions imply that Assumption 1 holds. \qed
\end{remark}

The \textit{key problem} is that we do not observe $U$. 
Therefore we can only identify the ``short'' conditional expectation of $Y$ given $D$ and $X$, i.e.
$$
g_s(D, X) := \Ep [Y \mid D, X].
$$
With the short regression in hand, we can compute proxies (or approximations) $\theta_s$ for $\theta$.
In particular, for the ATE and ACD the short parameter consists of $$\theta_s = \Ep [g_s(1,X) - g_s(0,X)] \text{ and } \theta_s = \Ep [\partial_d g_s(D,X)].$$

In our general framework, the proxy parameter
can also be expressed as the same linear functional applied to the short regression, $g_s(W_s)$ and the short regression is allowed to only be a mean-squared-projection of the CEF $\Ep[Y\mid D,X]$ onto a closed linear subspace $\Gamma_s$.

\begin{assumption}[\textbf{Proxy ``Short'' Parameter}] 
\label{asm:thetas}
The proxy parameter $\theta_s$ is defined by replacing the long regression $g$ with the short regression $g_s$ in the definition of the target parameter:
$$
\theta_s := \Ep m(W, g_s),
$$
where $g_s$ is the projection of the CEF $\Ep[Y\mid W_s]$ onto a closed linear subspace $\Gamma_s \subseteq \Gamma \cap L^2(P_{W_s})$, i.e.,
\begin{align}
    g_s :=~& \arg\min_{\gamma_s \in \Gamma_s} \Ep(Y-\gamma_s(W_s))^2 \equiv \arg\min_{\gamma_s \in \Gamma_s} \Ep (\Ep[Y\mid W_s] -\gamma_s(W_s))^2
\end{align}
We require $m(W, g_s)$ to be a measurable function of $W_s$ alone, i.e., the score depends only on $W_s$ when evaluated at the short regression $g_s$. Given this assumption, we use the shorthand notation $m(W_s, g_s)$ in place of $m(W, g_s)$. 
\end{assumption}

\noindent In the two working examples this assumption is satisfied, since $m(W_s, g_s)= g_s(1, X) - g_s(0, X)$ for the ATE and 
$m(W_s, g_s) = \partial_d g_s(D, X)$ for the ACD.

\subsection{Omitted variable bias for linear functionals}
\label{sec:ovb-general}

We now characterize the omitted variable bias, i.e., the difference between the ``long'' and ``short'' functionals, $\theta - \theta_s$, for general linear functionals of (projections of) the CEF. A key step to our approach is the following lemma that characterizes the target parameters and their proxies as inner products of regression functions with terms called Riesz representers (RR).

\begin{Lemma}[\textbf{Riesz Representation}]\label{Lemma:RR} Consider the long and short parameters $\theta$ and $\theta_s$ as given by Assumptions~\ref{asm:theta}~and~\ref{asm:thetas}.  There exist unique functions $\alpha \in \Gamma$ and $\alpha_s\in \Gamma_s$, referred to as the long and short Riesz Representers (RRs), such that 
$$
\Ep m(W, \gamma) = \Ep \gamma(W) \alpha(W), \quad\quad\quad \Ep m(W_s, \gamma_s) = \Ep \gamma_s(W_s) \alpha_s(W_s), 
$$
for all $\gamma\in \Gamma$ and $\gamma_s\in \Gamma_s$. 
Furthermore, $\alpha_s$ is the projection of $\alpha$ onto $\Gamma_s$, namely,
$$
\alpha_s \in \arg\min_{\gamma_s \in \Gamma_s} \Ep (\alpha(W) - \gamma_s(W_s))^2
$$
and therefore we also have that:
$$
\Ep (\alpha(W) - \alpha_s(W_s))^2 = \Ep \alpha^2 - \Ep \alpha^2_s.
$$
\end{Lemma}

\noindent An important aspect of Lemma~\ref{Lemma:RR} is the property that the short RR $\alpha_s$ is the orthogonal projection of the long RR $\alpha$ onto $\Gamma_s$. This property is crucial for the proof of our main omitted variable bias formula later in this section. Note that when $\Gamma, \Gamma_s$ are un-restricted spaces, i.e., $\Gamma=L^2(P_W)$ and $\Gamma_s=L^2(P_{W_s})$, then the latter property between the long and short RRs translates to:
$$
\alpha_s(W_s) = \Ep[ \alpha(W) \mid W_s].
$$
For restricted spaces, this is a more subtle and relatively surprising statement.

Intuitively, the Riesz representers can be seen as weights that map the outcome to the target parameter. In the case of the ATE with a binary treatment, and imposing no restrictions on the regression function, then the representers are just the classical inverse probability of treatment (Horvitz-Thompson) weights:
$$
\alpha(W) = \frac{D}{P(D=1\mid X,U)}-\frac{1-D}{P(D=0\mid X,U)}, \quad
\alpha_s(W) = \frac{D}{P(D=1\mid X)}-\frac{1-D}{P(D=0\mid X)}.
$$
This follows from change of measure arguments. One can also verify that  $\alpha_s = \Ep[\alpha\mid D, X]$,  by applying Bayes' rule. 

In the case of the ACD with a continuous treatment, using integration by parts we can verify that the representers are logarithmic derivatives of the conditional densities:
$$
\alpha(W) = -\partial_d \log f(D \mid X,U ), \quad
\alpha_s(W_s) = -\partial_d \log f(D \mid X).
$$

If we further restrict the regression functions to be partially linear as in Section~\ref{sec:plm} 
$$
g(W) = \beta D + f(X,U), \\
\quad g_s(W_s) = \beta_s D + f_s(X),
$$
then the restricted Riesz representers of either the ATE or the ACD functionals take the form:
$$
\alpha(W) = \frac{D- \Ep[D\mid X,U]}{\Ep (D- \Ep[D\mid X,U ])^2 }, \quad \alpha_s(W_s) = \frac{D- \Ep[D\mid X]}{\Ep (D- \Ep[D\mid X])^2 }.
$$
That is, the representer is given by the (scaled) residualized treatment, which we previously derived using the classical Frisch-Waugh-Lovell theorem, without invoking Riesz representation per~se. We remark here that Lemma~\ref{Lemma:RR} proves that the short RR in this case is the orthogonal projection of the long RR onto the space of functions of $D, X$ that are partially linear in $D$. This is a highly non-trivial statement that comes out of our general framework. For illustrative purposes, in the Online Appendix we provide a proof of this fact from first principles for this special case.

Using these lemmas, we obtain the following characterization of the OVB for general linear functionals of (projections of) the CEF. 

\begin{Theorem}[\textbf{OVB---General}]
\label{thm:ovb-general} 
Consider the long and short parameters $\theta$ and $\theta_s$ as given by Assumptions~\ref{asm:theta}~and~\ref{asm:thetas}. Then the OVB is given by the covariance between the regression error and the RR error,
$$
\theta - \theta_s 
= \Ep (g -g_s) (\alpha - \alpha_s ),
$$
with $\alpha,\alpha_s$ as defined in Lemma~\ref{Lemma:RR}. This bias can be further reparameterized in terms of $R^2$ measures, 
$$
\theta - \theta_s = \underbracket{
\frac{\Ep[(g - g_s)(\alpha - \alpha_s)] }{\sqrt{\Ep(g - g_s)^2\Ep(\alpha - \alpha_s)^2} } 
 }_{\rho} 
\times \underbracket{\sqrt{R^2_{ Y- g_s \sim g- g_s}}}_{C_Y}\times \underbracket{\sqrt{ \frac{(1-R^2_{\alpha\sim \alpha_s})}{R^2_{\alpha\sim \alpha_s}}}}_{C_D}\times \underbracket{\sqrt{\Ep(Y-g_s)^2 \Ep\alpha_s^2}}_{S}.
$$
\end{Theorem}  

This is the main conceptual result of the paper, and it is new. It covers a rich variety of causal estimands of interest, as long as they can be written as linear functionals of the long regression.  The Online Appendix verifies the OVB validity for many interesting examples beyond the running examples of the ATE and ACD in the main text. These include: (i)~weighted averages of potential outcomes~(Example~\ref{ex:wpo}); (ii)~weighted average treatment effects, such as the average treatment effect on the treated (ATT), or the average value of a policy~(Example~\ref{ex:att}); (iii)~average policy effects when transporting the short covariates~$W_s$~(Example~\ref{ex:policy}); (iv)~weighted average incremental effects~(Example~\ref{ex:waie}); and, (v)~policy effects from changing the distribution~of~$W_s$~(Example~\ref{ex:pecd}). 

\begin{remark}[Restricted vs Unrestricted Riesz Representers]
When the linear functional is assumed to be continuous for all $\gamma \in L^2(P_W)$ and not just for $\gamma\in \Gamma$, then the functional also admits short and long RRs $\bar\alpha\in L^2(P_W)$ and $\bar \alpha_s\in L^2(P_{W_s})$, that represent the functional for all square-integrable functions. The RR $\bar \alpha$ and $\bar\alpha_s$ do not necessarily coincide with their restricted counterparts $\alpha, \alpha_s$. Moreover, it is possible that $\alpha$ and $\alpha_s$ exist but $\bar \alpha$ and $\bar\alpha_s$ do not exist if it happens that the linear functional is continuous only on the subspace $\Gamma$. Whenever the un-restricted RRs exist then we show here that they satisfy the following relationship with the restricted RRs.

\begin{Lemma}\label{lemma:RR restricted}
Suppose that the linear functional $f\to \Ep m(W, f)$ is continuous for all $f\in L^2(P_W)$. Then there exists unique $\bar \alpha\in L^2(P_W)$ and $\bar\alpha_s\in L^2(P_{W_s})$, such that for all $f \in L^2(P_W)$ and all $f_s\in L^2(P_{W_s})$:
$$
\Ep m(W, f) = \Ep f(W) \bar\alpha(W), \quad\quad\quad \Ep m(W_s, f_s) = \Ep f_s(W_s) \bar\alpha_s(W_s), 
$$
Moreover, the restricted RRs $\alpha\in \Gamma$ and $\alpha_s\in \Gamma_s$ are given by the orthogonal projections of $\bar \alpha$ and $\bar \alpha_s$ on $\Gamma$ and $\Gamma_s$, respectively. Since projections reduce the norm, $\Ep \alpha^2 \leq \Ep \bar \alpha^2$ and $\Ep \alpha_s^2 \leq \Ep  \bar \alpha_s^2$. \qed 
\end{Lemma}
\end{remark}

\subsection{Making sense of the OVB formula} 
\label{sec:making-sense-general}
Theorem~\ref{thm:ovb-general} generalizes the warm-up result to fully nonlinear models, and general target parameters defined as linear functionals of the long regression. Though more abstract, the same discussions of Sections~\ref{sec:making-sense-plm}~and~\ref{sec:bound-plm} apply.  As before, the bias decomposes into an identified scaling factor $S$, and an unidentified bias factor $\rho C_YC_D$. The terms $C_Y$ and $C_D$ capture the components of the bias factor that measure the capacity of the omitted variables in generate errors in the outcome regression and in the Riesz representer. In particular, $R^2_{ Y- g_s \sim g- g_s}$  in the first factor measures the proportion of residual variation in the outcome explained by confounders; whereas $1-R^2_{\alpha \sim \alpha_s}$ in $C_D$ measures the proportion of  variation of the long RR generated by latent confounders. These components capture the maximum potential of $U$ to create bias, which is modulated by how systematically these errors are related, as given~by~$\rho\in[-1,1]$.

When we take $g=\Ep[Y\mid D, X,U]$ and $g_s=\Ep[Y\mid D,X]$ we have the same useful interpretation of $R^2_{ Y- g_s \sim g- g_s} = \eta^2_{Y \sim U\mid D, X}$. The interpretation of~$1-R^2_{\alpha \sim \alpha_s}$ and $C^2_D = (1-R^2_{\alpha \sim \alpha_s})/R^2_{\alpha \sim \alpha_s}$ can be further specialized for different estimands, as follows. 

\begin{remark}[\textbf{Interpretation of OVB for the ATE with a Binary Treatment}] 
\label{rmk:ate}
For the ATE, 
\begin{equation}
1-R^2_{\alpha_{\text{ATE}}\sim\alpha_{\text{ATE},s}} = 1-\frac{\Ep [1/\Var(D\mid X)]}{\Ep [1/\Var(D \mid X,U)]}, 
\qquad C^{2, \text{ATE}}_D = \frac{\Ep [1/\Var(D\mid X,U)]}{\Ep [1/\Var(D\mid X)]}-1. 
\end{equation}

That is, the OVB of the ATE depends on the relative gain in the average precision of the treatment model due to~$U$.\footnote{Precision is the inverse of the variance.}  Thus, the interpretation of $1-R^2_{\alpha\sim\alpha_s}$ for the ATE with a binary treatment parallels that of the partially linear model---compare it to equation~(\ref{eq:partialr2}). The sole distinction being that, here, gains in predictive power are measured by the relative increase in precision rather than the relative decrease in variance.\footnote{ 
This connection can be strengthened by considering a latent Gaussian confounder model $D =1 (D^*>0)$, where $D^* = g(X) - \mu U - \sqrt{1- \mu^2} \epsilon_D,$  with $\epsilon_D$ and  $U$ both mutually independent standard Gaussian, and also  independent of~$X$. Note that 
$g(X)$ is identified from the relation $\Ep[D \mid X] = \Phi (g(X))$.  Then $P[D=1 \mid X,U] = \Phi ( (g(X) - \mu U)/\sqrt{1-\mu^2})$, and $P[D=1 \mid X] = \Phi (g(X))$, from which the relative gain in precision can be computed. Then here the gain in precision is a monotone function of $\mu^2 =\eta^2_{D^*\sim U | X}$, the $R^2$ in the latent regression of $D^*$ on $U$, after adjusting for $X$. This connection may be useful for empirical work. 
} 
\qed
\end{remark}

\begin{remark}[\textbf{Interpretation of OVB for the ATT with a Binary Treatment}] 
\label{rmk:att}
For the ATT, 
\begin{equation}
1-R^2_{\alpha_{\text{ATT} }\sim \alpha_{\text{ATT},s}} = 1 -\frac{\Ep [\Odds(D\mid X)]}{\Ep [\Odds(D\mid X,U)]}, 
\qquad C^{2, \text{ATT}}_{D}=\frac{\Ep[\Odds(D \mid X,U)]}{\Ep[\Odds(D \mid X)]} -1. 
\end{equation}
That is, the OVB of the ATT depends on the relative gain in the \emph{average odds} of treatment due to~$U$. In particular, note the similarities with the traditional sensitivity parameter of Rosenbaum's sensitivity model \citep{rosenbaum:b1987,yadlowsky2022bounds} or marginal sensitivity models \citep{tan:jasa2006,dorn2023sharp}. We elaborate on this point further below in Section~\ref{sec:msm}. \qed
\end{remark}


\begin{remark}[\textbf{Interpretation of OVB for Average Causal Derivatives}] For the ACD example,
\begin{equation}
1-R^2_{\alpha_{\text{ACD}} \sim \alpha_{\text{ACD},s}} =  1- \frac{\Ep [(\partial_d \log f(D \mid X))^2]}{
\Ep [(\partial_d \log f(D \mid X,U))^2]}, 
\qquad 
C^{\text{2, ACD}}_D =\frac{\Ep [(\partial_d \log f(D \mid X,U))^2]}{
\Ep [(\partial_d \log f(D \mid X))^2]} -1,
\end{equation}
which can be interpreted as the relative {gain in information that the confounder $U$ provides about the location of $D$.}  Furthermore, if $D$ is homoscedastic Gaussian, conditional on both $X$ and $(X,U)$, we then have
$$
\partial_d \log f(D \mid X,U) = -\frac{D- \Ep[D \mid X,U]}{\Ep (D- \Ep[D \mid X,U])^2}, \quad \partial_d \log f(D \mid X) = -\frac{D- \Ep[D \mid X]}{\Ep (D- \Ep[D \mid X])^2},
$$
so that $1-R^2_{\alpha \sim \alpha_s}$ simplifies to the nonparametric $R^2$ of the latent variable with the treatment, similarly to the partially linear model, i.e, $1-R^2_{\alpha \sim \alpha_s}= \eta^2_{D\sim U\mid X}$.
\label{rmk:acd}
\qed
\end{remark}

\subsection{Relationship with marginal sensitivity models.}
\label{sec:msm}
Given their popularity and importance, here we expand on the differences between our omitted variable bias approach for sensitivity analysis, and sensitivity models based on worst-case assumptions on the change in odds-ratios, such as the ``marginal sensitivity model'' (MSM)  \citep{tan:jasa2006,dorn2023sharp,dorn2025doubly}. As these approaches usually restrict $D$ to be binary, we focus on this case, with the understanding that this is not necessary for our approach. We note that similar discussion applies to Rosenbaum's sensitivity model \citep{rosenbaum:b1987,yadlowsky2022bounds}.  

\subsubsection{Omitted variable bias, sensitivity analysis and worst-case assumptions.}

Traditional approaches to sensitivity analysis, such as the marginal sensitivity model (MSM), often start with some assumption about the strength  of $U$, and then determine how this assumption constrains the target parameter of interest, such as the average potential outcome, the ATE or the ATT.  For example, the MSM posits that unmeasured confounders $U$ can change the odds of treatment by \emph{at most} $\Lambda \ge 1$, for any value of the observed covariates $X$, i.e., 
$$
\Lambda^{-1}  \leq \frac{\Odds(D\mid X=x,U=u)}{\Odds(D\mid X=x)} \leq \Lambda, \qquad \forall x \in \mathcal{X},  u \in \mathcal{U}.
$$
Given this constraint, one derives bounds on various causal effects. See, for example, \cite{tan:jasa2006}, and more recently \cite{dorn2023sharp} and \cite{dorn2025doubly} for sharp bounds under the MSM.

The omitted variable bias approach we take here inverts this logic entirely. Instead of starting from a primitive assumption about confounder strength, and then deriving its logical implications, we ask a targeted question: what exactly must be known about the confounder to characterize the bias of a given estimand? By dissecting the anatomy of the bias, the OVB formula reveals how confounders must be restricted to bound that particular target estimand---and it also shows that different estimands have different sensitivities to unmeasured confounding. 

Concretely, for example, if one wants to bound the bias of the ATE, the OVB formula shows that it suffices to restrict the gains in \emph{average} precision of treatment assignment (Remark~\ref{rmk:ate}), whereas if one wants to bound the ATT, it suffices to restrict the ratio of \emph{average} odds of treatment (Remark~\ref{rmk:att}). If, instead, interest lies in a partially linear projection coefficient, it suffices to restrict the gains in variation explained of the treatment (Section~\ref{sec:plm}). Note these sensitivity parameters are not arbitrary; rather, they emerge naturally from the bias structure. Moreover, any restrictions beyond those are stronger than what is needed to bound each of these estimands.  

The OVB analysis thus reveals the marginal sensitivity model imposes stronger assumptions than what is needed to bound the bias of any of these estimands. The case of the ATT is particularly instructive. In that case, the sensitivity parameters of the MSM and the OVB formula are very similar in nature.  The key distinction is simply that the MSM asks researchers to make a strong {worst-case} assumption (i.e., bounding the worst-case change in odds) while the OVB formula shows that it suffices to make a weaker, average-case assumption (i.e., bounding the change in average odds). Understanding this distinction is important because $\Lambda$ can be unbounded even when $1-R^2_{\alpha \sim \alpha_s}$ is not only finite, but small (see Online Appendix for an example). 

\subsubsection{Using \texorpdfstring{$\Lambda$}{Lambda} to calibrate judgments on OVB parameters}

A researcher accustomed to the marginal sensitivity model might find this estimand-specific approach unfamiliar---after all, the appeal of the MSM lies partly in having a single parameter $\Lambda$ that can be used to bound various causal quantities. Theorem~\ref{thm:msm-ovb} bridges this gap, and shows how one can still leverage a single belief about the worst-case odds ratio $\Lambda$ to bound the maximum strength of $1-R^2_{\alpha \sim \alpha_s}$ for various estimands.\footnote{The proof of Theorem~\ref{thm:msm-ovb} leverages the fact that the long propensity score is bounded under the MSM. Thus, these results can potentially be extended to other sensitivity models that imply bounded propensity scores, such as Rosenbaum's model \citep{rosenbaum:b1987,yadlowsky2022bounds} or the model of \cite{masten:e2018}.} We state the theorem in terms of $C^2_D$ with the understanding that $1-R^2_{\alpha \sim \alpha_s} = C^2_D/(1+C^2_D)$.

\begin{Theorem}[Sharp Bounds on $C_D$ given $\Lambda$] 
\label{thm:msm-ovb}
Under the marginal sensitivity model we have,
\begin{alignat}{3}
C^{2,\text{ATT}}_{D}
&= \frac{\Ep [\Odds(D\mid X,U)]}{\Ep [\Odds(D\mid X)]} - 1 &
&\leq 
\frac{(\Lambda -1)^2}{\Lambda}
\times \left(1-\frac{P(D=1)}{\Ep[\Odds(D\mid X)]}\right)
&\leq \frac{(\Lambda -1 )^2}{\Lambda},
\nonumber\\
C^{2,\text{ATE}}_{D}
&= \frac{\Ep [1/\Var(D\mid X,U)]}{\Ep [1/\Var(D\mid X)]} -1 &
&\leq 
\frac{(\Lambda -1)^2}{\Lambda}
\times \left(1-\frac{3}{\Ep[1/\Var(D\mid X)]}\right)
&\leq \frac{(\Lambda -1 )^2}{\Lambda},
\nonumber\\
C^{2, \text{PLM}}_{D}
&= \frac{\Ep [\Var(D\mid X)]}{\Ep [\Var(D\mid X,U)]} -1&
&\leq  \frac{(\Lambda - 1)^2}{\Lambda}\times \left(\frac{\Ep\left[\omega(X) \times \Var(D\mid X)\right]}{\Ep[\omega(X)]}\right)
&\leq \frac{(\Lambda -1 )^2}{4\Lambda},
\nonumber
\end{alignat}
where $\omega(X):= \Var(D\mid X)/(\Lambda  + (\Lambda -1)^2\Var(D\mid X))$. The first bound is sharp given $\Lambda$ and the observed data. The second bound is sharp given $\Lambda$ alone.
\end{Theorem}

How could one use these relationships?  If a researcher truly believes the {worst-case odds} ratio to be at most $\Lambda$, then they should use the recent sharp bounds of \cite{dorn2023sharp} and \cite{dorn2025doubly}. If, however, researchers find the worst-case assumption too strong, they can use $\Lambda$ as a useful \emph{heuristic} to calibrate plausible values for  the OVB parameter $1-R^2_{\alpha \sim \alpha_s}$ from first principles, across various estimands, while effectively relaxing the strict worst-case assumption of the MSM.

\subsection{Using observed covariates to calibrate the strength of \texorpdfstring{$U$}{U}.}

Beyond making direct plausibility judgments on the strength of confounding using the above quantities, analysts can also leverage judgments of relative importance of variables to bound the size of the bias (see, e.g. \citealp{imbens2003sensitivity,CH2020}). For instance, if one has reasons to believe that $U$ would not generate as much gains in explanatory power as certain key observed covariates $X_j$, this can be used to formally place bounds on the strength of confounding due to $U$. This allows one to assess, for instance, whether confounders as strong or stronger than observed covariates would be sufficient to overturn an empirical result. We elaborate the benchmarking procedure formally in the Online Appendix and illustrate its use in the empirical example. These results extend previous benchmarking ideas for linear regression models to the general case.

\section{Statistical Inference Under Omitted Variable Bias}
\label{sec:inference}

Suppose the components $\rho$, $C_Y$, $C_D$ of the OVB formula in Theorem~\ref{thm:ovb-general} are set through hypotheses on the maximum explanatory power of omitted variables. We then have the following  bounds $[\theta_{-}, \theta_+]$ for the target parameter $\theta$: 
\begin{equation}
\label{eq:bounds}
\theta_{\pm} =\theta_s \pm |\rho| C_Y C_D S, \quad S^2 = \Ep (Y-g_s)^2\Ep \alpha^2_s.
\end{equation}

The estimable components of the bounds are $S$ and $\theta_s$.  We can estimate these components via debiased machine learning (DML), which is a form of the classical ``one-step'' semi-parametric correction \citep{levit1975efficiency,hasminskii1978,pfanzagl:wefelmeyer, bickel1993efficient,newey1994,DML,DML-LR} based on Neyman orthogonal scores we give for these components, combined with cross-fitting, an efficient form of data-splitting.

For debiased machine learning of $\theta_s$, we exploit the representation
$$
\theta_s = \Ep [m(W_s, g_s) + (Y- g_s) \alpha_s],
$$
as in \cite{AutoDML-Lasso, AutoDML}.  This representation is Neyman orthogonal with respect to  perturbations of $(g_s, \alpha_s)$, which is a key property required for DML.\footnote{For the partial linear model of Section~\ref{sec:plm}, one can alternatively use the partialling-out approach of \cite{robinson}.}  Another component to be estimated is
$$
\Ep (Y-g_s)^2 =: \sigma^2_{s},
$$
which is also Neyman-orthogonal with respect to $g_s$.  The final component to be estimated is $\Ep \alpha^2_s$. For this we explore the following formulation:
$$
\Ep \alpha^2_s = 2 \Ep m(W_s, \alpha_s) - \Ep \alpha^2_s =: \nu^2_s,
$$
where the latter parameterization is  Neyman-orthogonal. Specifically Neyman orthogonality refers to the property:
\begin{eqnarray*}
&& \partial_{g,\alpha} \Ep [m(W_s, g) + (Y- g) \alpha] \Big |_{\alpha_s, g_s} \!\!\!=0; \ \partial_{g} \Ep (Y-g)^2 \Big |_{g_s} \!\!\!= 0; \ \partial_{\alpha}   \Ep [2 m(W_s, \alpha) -  \alpha^2] \Big |_{ \alpha_s} \!\!\!= 0;
\end{eqnarray*}
where $\partial$ is the Gateaux (pathwise derivative) operator over directions $h \in \Gamma_s$. These Neyman orthogonality properties follow from the definition of $g_s$ as the minimizer over a closed linear space $\Gamma_s$, the representation property of $\alpha_s$ over $\Gamma_s$ and the fact that $g_s, \alpha_s, h\in \Gamma_s$, which imply the conditions:
\begin{align*}
    \forall \gamma_s\in \Gamma_s: \Ep (Y - g_s(W_s)) \gamma_s(W_s)=~&0,
    &
    \Ep [m(W_s,\gamma_s) - \alpha_s(W_s) \gamma_s(W_s)]=~&0.
\end{align*}

Application of DML theory in \cite{DML} and the delta-method gives the statistical properties of the estimated bounds under the condition that machine learning of  $g_s$ and $\alpha_s$ is of sufficiently high quality, with learning rate faster than $n^{-1/4}$. The estimation relies on the following generic algorithm.
\begin{definition}[DML($\psi$)] \label{alg:target} Input the Neyman-orthogonal score $\psi(Z; \beta, \eta)$, where $\eta =  (g, \alpha)$. Then (1), given a sample $(Z_i:=(Y_i,D_i, X_i))_{i=1}^n$, randomly partition the sample into folds $(I_{\ell})_{\ell=1}^L$ of approximately equal size. Denote by $I_{\ell}^c$ the complement of $I_{\ell}$. (2) For each $\ell$, estimate $\widehat \eta_\ell = (\widehat{g}_{\ell}, \widehat{\alpha}_{\ell})$ from observations in $I_{\ell}^c$. (3) Estimate $\beta$ as a root of:
    $
  0=n^{-1}\sum_{\ell=1}^L\sum_{i\in I_{\ell}} \psi(\beta, Z_i; \widehat \eta_\ell).$
 Output $\widehat \beta$ and the estimated scores $\widehat \psi^o (Z_i) =\psi(\widehat \beta, Z_i; \widehat \eta_{\ell})$ for each $i \in I_\ell$ and each $\ell$.
\end{definition}

Therefore the estimators are defined as 
$$
\widehat \theta_s := \mathrm{DML} (\psi_{\theta_s});  \quad \widehat \sigma^2_s := \mathrm{DML} ( \psi_{\sigma_s^2});  \quad \widehat \nu^2_s := \mathrm{DML} (\psi_{\nu_s^2});
$$
for the scores
\begin{eqnarray*}
&& \psi_{\theta_s} (Z; \theta_s, g_s, \alpha_s) := m(W_s, g_s) + (Y- g_s(W_s)) \alpha_s(W_s) - \theta_s; \\
&& \psi_{\sigma_s^2}(Z; \sigma^2_s, g_s)  :=  (Y- g_s(W_s))^2 - \sigma_s^2; \\
&& \psi_{\nu_s^2} (Z; \nu^2_s, \alpha_s) := (2 m(W_s, \alpha_s) - \alpha_s^2) - \nu_s^2.
\end{eqnarray*}

We say that an estimator $\hat \beta$ of $\beta$ is asymptotically linear and Gaussian with the centered influence function $\psi^o(Z)$ if
$$
\sqrt{n} (\hat \beta - \beta) = \frac{1}{\sqrt{n}} \sum_{i=1}^n \psi^o(Z_i) + o_{\Pr} (1) \leadsto N(0, \Ep \psi^{o2}(Z)).
$$

The application of the results in \cite{DML} for linear score functions yields the following result.

\begin{Lemma}[\textbf{DML for Bound Components}]\label{lemma:DML} Suppose that each of $\psi$'s listed above and the machine learners $\hat \eta_\ell = (\hat{\alpha}_\ell, \hat{g}_\ell)$ of $\eta_0 = (g_s, \alpha_s)$ in $L^2(P_{W_s})$ obey Assumptions 3.1 and 3.2 in \cite{DML}, in particular the rate of learning $\eta_0$ in the $L^2(P_{W_s})$ norm needs to be $o_P(n^{-1/4})$. Then the estimators are asymptotically linear
and Gaussian with influence functions:
$$
\psi^o_{\theta_s}(Z)  : = \psi_{\theta_s} (Z; \theta_s, g_s, \alpha_s); \quad \psi^o_{\sigma_s^2}(Z) := \psi_{\sigma_s^2}(Z; \sigma^2_s, g_s); \quad \psi^o_{\nu_s^2} (Z) := \psi_{\nu_s^2} (Z; \nu^2_s, \alpha_s).
$$
The covariance of the scores can be estimated by the empirical analogues using the covariance of the estimated scores.
\end{Lemma}

The resulting plug-in estimator for the bounds is then:
$$
\widehat \theta_{\pm} = \widehat \theta_s \pm    |\rho| C_Y C_D \widehat S, \quad \widehat S^2 = \widehat \sigma^2_s \widehat \nu^2_s.
$$
Confidence bounds for the bounds can be constructed using the following result.

\begin{Theorem}[\textbf{DML Confidence Bounds for Bounds}]
\label{thm:DML} Under the conditions of Lemma~\ref{lemma:DML}, the plug-in estimator $\widehat \theta_{\pm}$ is also asymptotically linear and Gaussian with the influence function:
$$
\varphi^o_{\pm}(Z) = \psi^o_{\theta_s} (Z) \pm \frac{|\rho|}{2} \frac{C_Y C_D}{S} 
( \sigma^2_s \psi^o_{\nu_s^2}(Z) + \nu_{s}^2 \psi^o_{\sigma_s^2}(Z)), \quad S>0.
$$
Therefore, the confidence bound
$$
[\ell, u] = \left [\widehat \theta_{-}  - \Phi^{-1}(1- a) \sqrt{\frac{\Ep \varphi^{o2}_{-}}{ n}}, \  \widehat \theta_+ + \Phi^{-1}(1- a) \sqrt{\frac{\Ep \varphi^{o2}_{+}}{n}} \right]
$$
has the one-sided covering property, namely
$$
\Pr( \theta_- \geq \ell) \to 1- a \text{ and } \Pr( \theta_+ \leq u) \to 1- a,~\text{as}~n\to \infty.
$$
The same results continue to hold if $\Ep \varphi^{o2}_{\pm}(Z)$  is replaced by the empirical analogue $
\frac{1}{n} \sum_{\ell=1}^L \sum_{i \in I_\ell} \hat \varphi^{o2}_{\pm} (Z_i).$
\end{Theorem}

We focus on the one-sided covering property stated in the theorem, since in applications the relevant hypotheses are typically one-sided. We can use  further adjustments of \cite{stoye:CI} to construct uniformly valid two-sided intervals (in broad terms, Stoye's approach smoothly switches to two-sided bands slightly before the bounds  cannot be distinguished from a singleton by any statistical test).

The following remark discusses learning the regression function~$g_s$ and the Riesz representer~$\alpha_s$.

\begin{remark}[\textbf{Machine Learning of $\alpha_s$ and $g_s$}]\label{remark:riesz}
Estimation of the short regression $g_s$ is standard and a variety of modern methods can be used (neural networks, random forests, penalized regressions). Estimation of the short RR $\alpha_s$ can proceed in one of the following ways. First, we can use analytical formulas for $\alpha_s$ (see e.g.,  \cite{DML,semenova2021debiased}, and references therein, for practical details). Second, we can use a variational characterization of $\alpha_s$:
$$
\alpha_s = \arg\min_{\alpha \in \mathcal{A}} \Ep [  \alpha^2(W_s)- 2m(W_s, \alpha)], 
$$
where $\mathcal{A}\subseteq \Gamma_s$ is the parameter space for $\alpha_s$, as proposed in \cite{AutoDML,AutoDML-Lasso}. This avoids inverting propensity scores or conditional densities, as usually required when using  analytical formulas. 
This approach is motivated by the first-order-conditions of the  variational characterization: $$\Ep \alpha_s g = \Ep m(W_s, g) \quad \text{  for all $g$ in $\mathcal{G}$, } $$
which is the definition of the RR.  Neural network (RieszNet) and random forest (ForestRiesz) implementations of this approach are given in \cite{riesznet}, and the Lasso implementation in  \cite{AutoDML-Lasso}.\footnote{ A third option is to use a minimax (adversarial) characterization of $\alpha_s$, as in \cite{RR-local,Adverse}: 
\mbox{$
\alpha_s = \arg\min_{\alpha \in \mathcal{A}} \max_{g \in \mathcal{G}} | \Ep m(W_s, g) - \Ep \alpha g |,
$}
where $\mathcal{A}$ is the parameter space for $\alpha_s$.
The Dantzig selector implementation of this approach is given in \cite{RR-local}. The neural network implementation of this approach is given in \cite{Adverse}.}\qed
\end{remark}

\section{Omitted Firm Characteristics in Evaluating the Effects of 401(k) Plan.}\label{sec:empirical-examples}

In this section we demonstrate the utility of our approach in an empirical example that estimates the average treatment effect of 401(k) eligibility on net financial assets \citep{pvw:94, pvw:95,DML}.  Our goal is to determine whether prior conclusions, reached under the assumption of conditional ignorability, are robust to plausible scenarios of unmeasured  confounding.  This example illustrates sensitivity analysis for the ATE in a partially linear model and in a nonparametric model with a binary treatment. In the Online Appendix we provide an additional example that estimates the price elasticity of gasoline demand \citep{blundell2012measuring, blundell2017nonparametric, chetverikov2017nonparametric} and illustrates sensitivity analysis for the average causal derivative with a continuous treatment.

\subsection{Estimates under conditional ignorability.}

A 401(k) plan is an employed sponsored tax-deferred savings option that allows individuals to deduct  contributions from their taxable income, and accrue tax-free interest on investments within the plan.  Introduced in the early 1980s as an incentive to increase individual savings for retirement, an important question in the savings literature is precisely to quantify the \emph{causal} impact of 401(k) eligibility on net financial assets. Indeed, a naive comparison of net financial assets between those individuals with and without 401(k) eligibility suggests a positive and large impact: using data from the 1991 \emph{Survey of Income and Program Participation} (SIPP), this difference amounts to~\$19,559.

The problem of this naive comparison, however, is that 401(k) plans can be obtained only by those individuals that work for a firm that offers such savings option---and employment decisions are far from randomized. As an attempt to overcome this lack of random assignment, \cite{pvw:94}, \cite{pvw:95}, and more recently \cite{DML}, leveraged the 1991 SIPP data to adjust for potential confounding factors between 401(k) eligibility and the financial assets of an individual.  As explained in \cite{pvw:94}, at least around the time 401(k) plans initially became available, people were unlikely to make employment decisions based on whether an employer offered a 401(k) plan; instead, their main focus was on salary and other aspects of the job. Thus, as a first approximation, whether one is eligible for a 401(k) plan could be taken as ignorable once we condition on income and other covariates related to job choice.

\begin{figure}[p]
\begin{subfigure}[t]{0.5\linewidth}
\centering
\begin{tikzpicture} \large
\node (D) at (0,0) {$D$};
\node (X) at (2,-2) {$X$};
\node (Y) at (4,0) {$Y$};
\node[draw, circle, dashed] (F) at (0,-2) {$U$};
\node[draw, circle, dashed] (U) at (2,-3.5) {$L$};
\draw[thick, ->] (D) -- (Y);
\draw[thick, ->] (U) -- (F);
\draw[thick, ->] (U) -- (X);
\draw[thick, ->] (U) -- (D);
\draw[thick, ->] (X) -- (Y);
\draw[thick, ->] (F) -- (D);
\draw[thick, ->] (X) -- (D);
\end{tikzpicture}
\caption{\small Ignorability holds conditional on $X$ only.}
\label{fig:dag-401k-1}
\end{subfigure}%
\begin{subfigure}[t]{0.5\linewidth}
\centering
\begin{tikzpicture} \large
\node (D) at (0,0) {$D$};
\node (X) at (2,-2) {$X$};
\node (Y) at (4,0) {$Y$};
\node[draw, circle, dashed] (M) at (2.5,-.75) {$M$};
\node[draw, circle, dashed] (F) at (0,-2) {$U$};
\node[draw, circle, dashed] (U) at (2,-3.5) {$L$};
\draw[thick, ->] (D) -- (Y);
\draw[thick, ->] (U) -- (F);
\draw[thick, ->] (U) -- (X);
\draw[thick, ->] (U) -- (D);
\draw[thick, ->] (X) .. controls(3, -1.5) .. (Y);
\draw[thick, ->] (D) -- (M);
\draw[thick, ->] (F) -- (D);
\draw[thick, ->] (F) -- (M);
\draw[thick, ->] (X) -- (M);
\draw[thick, ->] (M) -- (Y);
\draw[thick, ->] (X) -- (D);
\end{tikzpicture} 
\caption{\small Ignorability holds conditional on $X$ \emph{and} $U$.}
\label{fig:dag-401k-2}
\end{subfigure}
\caption{ \small Two possible causal DAGs for the 401(k) example.}
\label{401K}
\vspace*{3in}
\end{figure}

It is useful to think about causal diagrams \citep{pearl:causality} that represent this identification strategy. One possible model is shown Figure~\ref{fig:dag-401k-1}.  Here the outcome variable, $Y$, consists of net financial assets;\footnote{Defined as the sum of IRA balances, 401(k) balances, checking accounts, U.S. saving bonds, other interest-earning accounts in banks and other financial institutions, other interest-earning assets (such as bonds held personally), stocks, and mutual funds less non-mortgage debt.} the treatment variable, $D$, is an indicator for being eligible to enroll in a 401(k) plan; finally, the vector of observed covariates, $X$, consists of: (i) age; (ii) income; (iii) family size; (iv) years of education; (iv) a binary variable indicating marital status; (v) a ``two-earner'' status indicator; (vi) an IRA participation indicator; and, (vii) a home ownership indicator.
We consider that the decision to work for a firm that offers a 401(k) plan depends both on the observed covariates~$X$, but also on \emph{unobserved} firm characteristics, denoted by~$U$; moreover, $X$, $U$, and $D$ are jointly affected by a set of latent factors $L$. Most importantly, note the assumption of \emph{absence} of direct arrows, both from $U$ and $L$, to $Y$. Under such assumption, conditional ignorability holds adjusting for $X$ only. The story represented by the DAG of Figure~\ref{fig:dag-401k-1} is one way of rationalizing the identification strategy used in earlier papers.

\begin{table}[p] 
\centering 
\begin{tabular}{lrrrrrrr}
\hline \hline
& \multicolumn{3}{c}{Results Under Conditional Ignorability} & & \multicolumn{1}{c}{Robustness Values}\\
\cmidrule{2-4} \cmidrule{6-6}
Model               & Short Estimate  & Std. Error    & Confidence Bounds & & $\text{RV}_{\theta =0,~a = 0.05}$ \\
\hline
Partially Linear    & 9,002           & 1,394         & [6,271; 11,733] & & 5.4\%               \\
Nonparametric       & 7,949           & 1,245         & [5,509; 10,388] & & 4.5\%               \\
\hline \hline
\end{tabular}
\caption{\small Minimal sensitivity reporting.  Significance level of 5\%.}
\label{tab:minimal}
\end{table}

\begin{table}
{\centering
\begin{tabular}{lcccc}
\hline
\hline
Model               & Short Estimate & |Bias| Bound & ATE Bounds & Confidence Bounds   \\ \hline
Partially Linear    & 9,002 (1,394) & 4,196 (316) & [4,808; 13,196]  & [2,497; 15,582]\\
    Nonparametric & 7,949 (1,245) & 4,516 (336) & [3,452; 12,460] & [1,383; 14,630]\\
\hline
\hline
\end{tabular}
} 
\caption{\small Estimate, bias, and bounds for the ATE.  Significance level of 5\%. Standard errors in parenthesis. Confounding scenario: $\rho^2 = 1$; $C^2_Y \approx 0.04$; $C^2_D\approx 0.03$.
}
\label{tab:401k}
\end{table}

\begin{figure}[p]
\begin{subfigure}[t]{0.5\linewidth}
\centering
    \includegraphics[scale=.65]{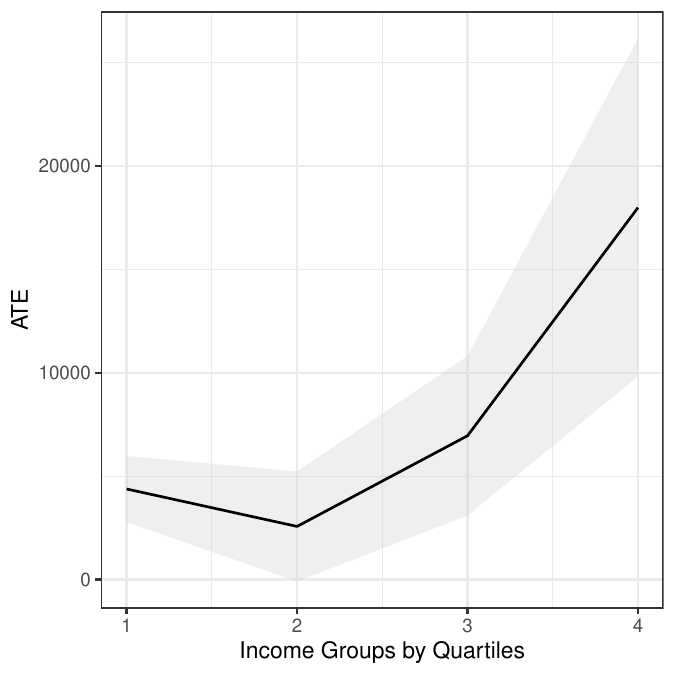}
    \caption{\small Estimates under no confounding.}
    \label{fig:401k-no-confounding}
\end{subfigure}%
\begin{subfigure}[t]{0.5\linewidth}
\centering
    \includegraphics[scale=.65]{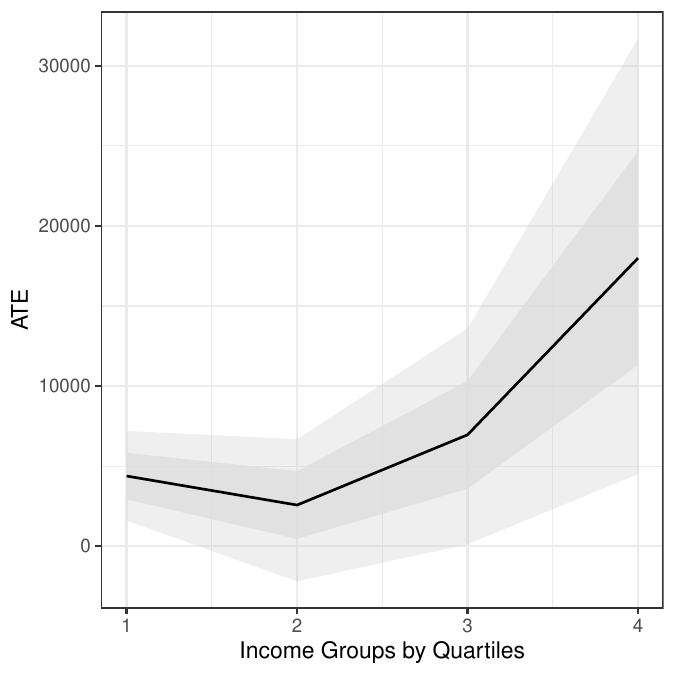}
    \caption{\small Bounds under posited confounding.}
    \label{fig:401k-confounding}
\end{subfigure}
\caption{\small Estimate (black), bounds (dark grey), and confidence bounds (light grey) for the ATE  by income quartiles.  Confounding scenario: $\rho^2 = 1$; $C^2_Y \approx 0.04$; $C^2_D\approx 0.03$. Significance level of $5\%$.} 
\label{fig:401K}
\vspace*{3in}
\end{figure}

The first three columns of Table~\ref{tab:minimal} shows the estimates for the average treatment effect  (ATE) of 401(k) eligibility on net financial assets under this conditional ignorability assumption. For these estimates, we follow the same strategy used in \cite{DML}, and we estimate the ATE using DML with Random Forests, considering both a partially linear model (PLM), and a nonparametric model (NPM).\footnote{We use Random Forest both for the outcome and treatment regression and estimate the parameters using DML with 5-fold cross-fitting. In order to reduce the variance that stems from sample splitting, we repeat the procedure 5 times. Estimates are then combined using the median as the final estimate, incorporating variation across experiments into the standard error as described in \cite{DML}. \label{foot:dml-fit}} As we can see, after flexibly taking into account observed confounding factors, although the estimates of the effect of 401(k) eligibility on net financial assets are substantially attenuated, they are still large, positive and statistically significant (approximately \$9,000 for the PLM and \$8,000 for the NPM).  With the nonparametric model, we further explore heterogeneous treatment effects, by analyzing the ATE within income quartile groups. The results are shown in Figure~\ref{fig:401k-no-confounding}. We see that the ATE varies  substantially across groups, with effects ranging from approximately \$4,000 (first quartile) to almost \$18,000 (last quartile).

\subsection{Sensitivity analysis}
It is now useful to consider scenarios in which conditional ignorability  fails. Figure~\ref{fig:dag-401k-2} presents one such scenario, where a violation of conditional ignorability is credible.\footnote{We note that Figure~\ref{fig:dag-401k-2} is just one example, and our sensitivity analysis results hold for {any} model in which conditional ignorability holds given observed variables and unobserved confounders.}  Employers often offer a benefit in which they ``match'' a proportion of an employee's contribution to their 401(k) up to  5\% of the employee's salaries. The model in Figure~\ref{fig:dag-401k-2} allows this  ``matched amount,'' denoted by $M$, to be determined by unobserved firm characteristics $U$, observed worker characteristics $X$, and by 401(k) eligibility $D$.
In this model, adjustment for $X$ alone is \emph{not} sufficient for control of confounding.  Instead, we now need to condition \emph{both} on observed covariates $X$ \emph{and} unobserved confounders $U$ for ignorability to hold. How strong would the omitted firm characteristics $U$ have to be in order to overturn our previous conclusions?  And how plausible are the strengths revealed to be problematic? In what follows, we use our sensitivity analysis results to address these questions.

\subsubsection{Minimal sensitivity reporting.}

In reporting empirical results, the following definition will be useful.  This definition extends the suggestion of \cite{CH2020, cinelli2020omitted} made for linear regression to the general case.
 
\begin{definition}[Robustness Values] The robustness value $\text{RV}_{\theta, a}$ stands  for the minimum upper bound $RV$ on  sensitivity parameters,
$R^2_{Y-g_s \sim g-g_s} \leq \text{RV}$ and \mbox{$1-R^2_{\alpha\sim \alpha_s}\leq \text{RV}$}, $|\rho| \leq 1$, such that the confidence bound $[l,u]$ of Theorem~\ref{thm:DML} includes $\theta$, at the significance level~$a$.
\end{definition}
Whereas standard errors,  t-values or p-values communicate how robust the short estimate is to \emph{sampling errors}, the idea of robustness values is to quickly communicate how robust the short estimate is to \emph{systematic errors} due to residual confounding.  For example, $\text{RV}_{\theta = 0,~a=.05}$ measures the minimal strength on both confounding factors  such that the estimated confidence bound for the ATE would include zero, at the 5\% significance level.\footnote{As put by \citet{rosenbaum2005sensitivity}, ``a sensitivity analysis [...] asks what the unmeasured covariate would have to be like to alter the conclusions of the study.'' Measures such as the RV provide such a characterization, and have a long history in the sensitivity  analysis literature, see, e.g. \cite{rosenbaum:b1987} and \cite{evalue}.} 

Table~\ref{tab:minimal} illustrates our proposal for a minimal sensitivity reporting of causal effect estimates. Beyond the usual estimates under the assumption of conditional ignorability, it reports the robustness values of the short estimate. Starting with the PLM, the $\text{RV}_{\theta = 0, a = 0.05}= 5.4\%$ means that unobserved confounders that explain less than 5.4\% of the residual variation, \emph{both} of the treatment, and of the outcome, are not sufficiently strong to bring the lower limit of the confidence bound to zero,  at the 5\% significance level. Moving to the  nonparametric model, we obtain a similar, but somewhat lower value of $\text{RV}_{\theta = 0, a = 0.05} = 4.5\%$. The RV thus provides a quick and meaningful reference point that summarizes the robustness of the short estimate against unobserved confounding---any postulated confounding scenario that does not meet this minimal criterion of strength cannot overturn the results of the original study. 

\subsubsection{Main confounding scenario.}   We now proceed to construct a particular confounding scenario, based on the contextual details of the problem. We start with the  assumption that $U$ explains as much variation in net financial assets as the total variation of the maximal matched amount of income (5\%) over the period of three years (roughly the period over which the effect is measured).\footnote{This strategy is based on a suggestion by James Poterba.} In the worst case scenario, this would lead to an additional $3\%$ of \emph{total} variation explained, resulting in a \emph{partial} $R^2$ of outcome with omitted firm characteristics $U$ of $C_Y^2 = \eta^2_{Y \sim U \mid  DX} = 4\%$.\footnote{
$
\eta^2_{Y \sim U \mid  DX} = \frac{\eta^2_{Y \sim UDX} - \eta^2_{Y \sim DX}}{1 - \eta^2_{Y \sim DX}}  = \frac{0.28 + 0.03 - 0.28}{1-0.28} \approx 4\%,
$
}
This amounts to a relative increase of approximately 10\%  in the baseline $R^2$ of the outcome regression of 28\%. Following similar reasoning, and more conservatively, we posit that omitted firm characteristics can explain an additional $2.5\%$ of the variation in 401(k) eligibility, corresponding to a $22\%$ \text{relative} increase in the baseline $R^2$ of the treatment regression of 11.4\%. For the partially linear model, this results in
$1-R^2_{\alpha \sim \alpha_s}= \eta^2_{D\sim U\mid X}\approx 3\%$ (and also $C^2_D \approx 3\%$).\footnote{$
1-R^2_{\alpha \sim \alpha_s} = \eta^2_{D \sim U \mid X}  =  \frac{\eta^2_{D \sim UX} - \eta^2_{D \sim X} } { 1- \eta^2_{D \sim X}}  = \frac{0.114 + .025 - 0.114}{1-0.114} \approx 3\%.
$} We adopt the same scenario for the nonparametric model, with the understanding that now this would correspond to gains in precision rather than gains in variation explained (see Remark~\ref{rmk:ate}). Since both $\eta^2_{Y \sim U \mid DX} \approx 4\%$ and $1-R^2_{\alpha \sim \alpha_s} \approx 3\%$  are below the robustness value of 5.4\% (or 4.5\%), we immediately conclude that such confounding scenario is \emph{not} capable of bringing the lower limit of the confidence bound of the ATE to zero.  

The exact bias, bounds, and confidence bounds on the ATE implied by the posited  scenario are shown in Table~\ref{tab:401k}.\footnote{We use the same estimation procedure as described in footnote~\ref{foot:dml-fit}.} Starting with the partially linear model, the confounding scenario has an estimated absolute value of the bias of ~$\$4,196$.  Accounting for statistical uncertainty, we obtain a lower limit for the confidence bound of~$\$2,497$. The results for the  nonparametric model are qualitatively similar, with a bias of similar magnitude, and point estimates, bounds, and confidence bounds for the ATE shifted down by roughly one thousand dollars. Confidence bounds for group-wise ATEs can also be computed, and are shown in Figure~\ref{fig:401k-confounding}. Note how the bounds are still largely positive, with only a small excursion into the negative side in the case of the second quartile group.  These results suggest that the main qualitative findings reported in earlier studies are relatively robust to plausible violations of unconfoundedness, such as the one specified by our confounding scenario. 

\subsubsection{Sensitivity contour plots and benchmarks}
A useful tool for visualizing the whole sensitivity range of the target parameter, under different assumptions regarding the strength of confounding, is a bivariate contour plot showing the collection of curves in the space of $R^2$ values along which the confidence bounds are constant \citep{imbens2003sensitivity, CH2020}. These plots allow investigators to quickly and easily assess the robustness of their findings against \emph{any} postulated  confounding scenario.  Here we focus on contour plots for the lower limit of the confidence bounds, as this is the direction of the bias that threatens the preferred hypothesis in this empirical example. Analogous contours can be constructed for the upper limit of the confidence bounds, and are omitted.

\begin{figure}[p]
\begin{subfigure}[t]{0.5\linewidth}
\centering
    \includegraphics[scale=.62]{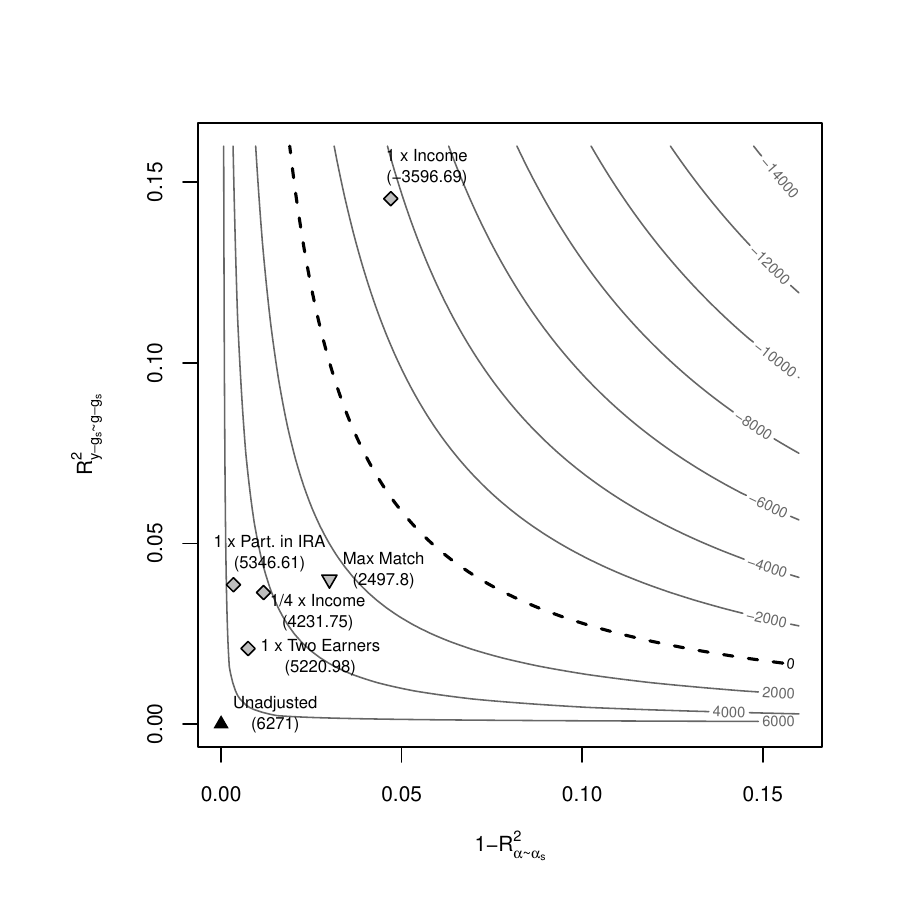}
    \caption{\small Lower limit confidence bound, PLM, $|\rho| =1$.}
    \label{fig:plm-lower}
\end{subfigure}%
\begin{subfigure}[t]{0.5\linewidth}
\centering
    \includegraphics[scale=.62]{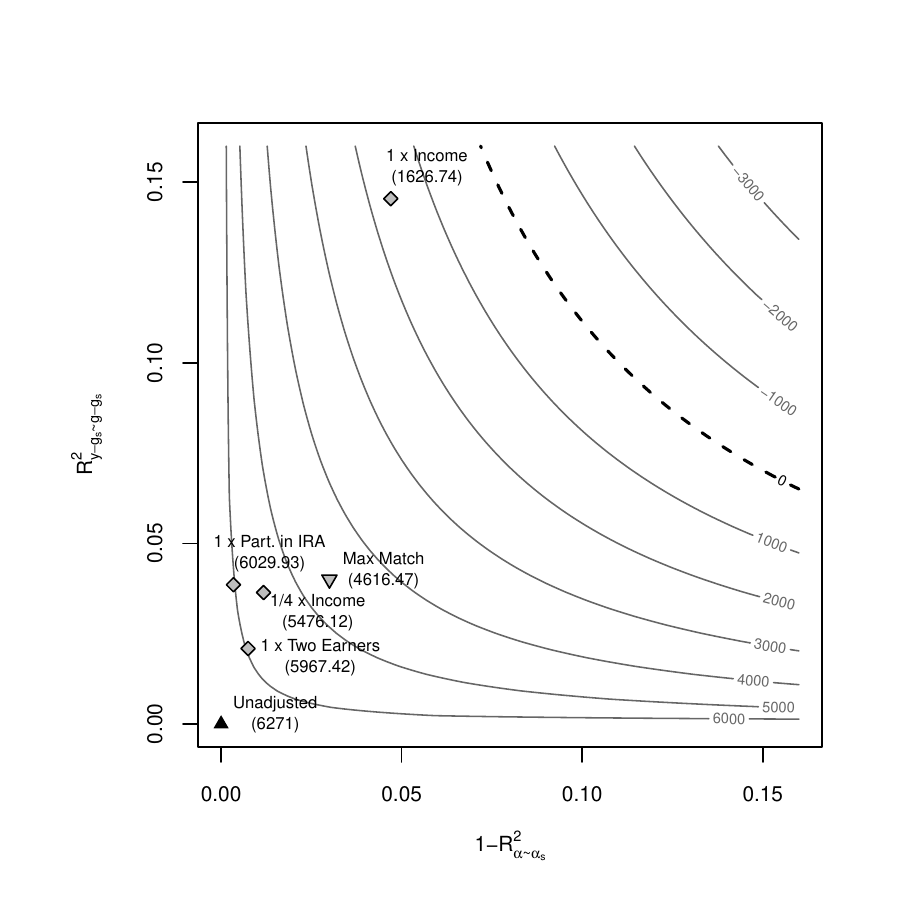}
    \caption{\small Lower limit confidence bound, PLM, $|\rho|=1/2$.}
    \label{fig:plm-lower-rho}
\end{subfigure}
\begin{subfigure}[t]{0.5\linewidth}
    \includegraphics[scale=.62]{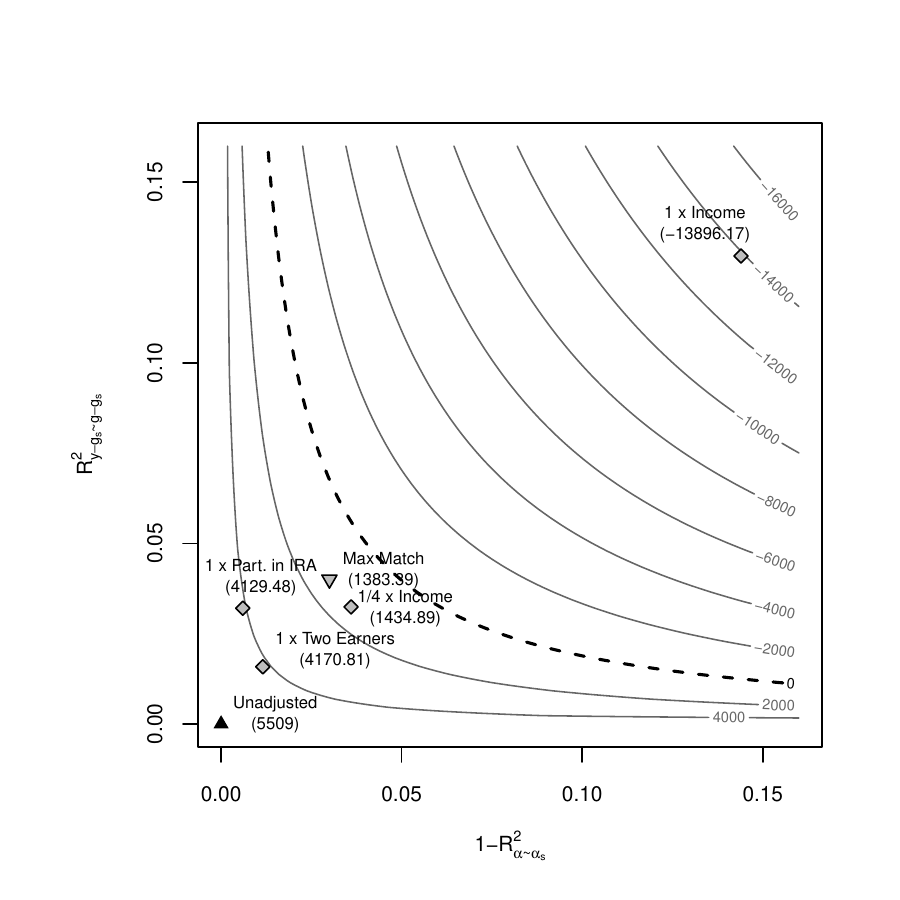}
    \caption{\small Lower limit confidence bound, NPM, $|\rho| =1$.}
    \label{fig:npm-lower}
\end{subfigure}%
\begin{subfigure}[t]{0.5\linewidth}
    \includegraphics[scale=.62]{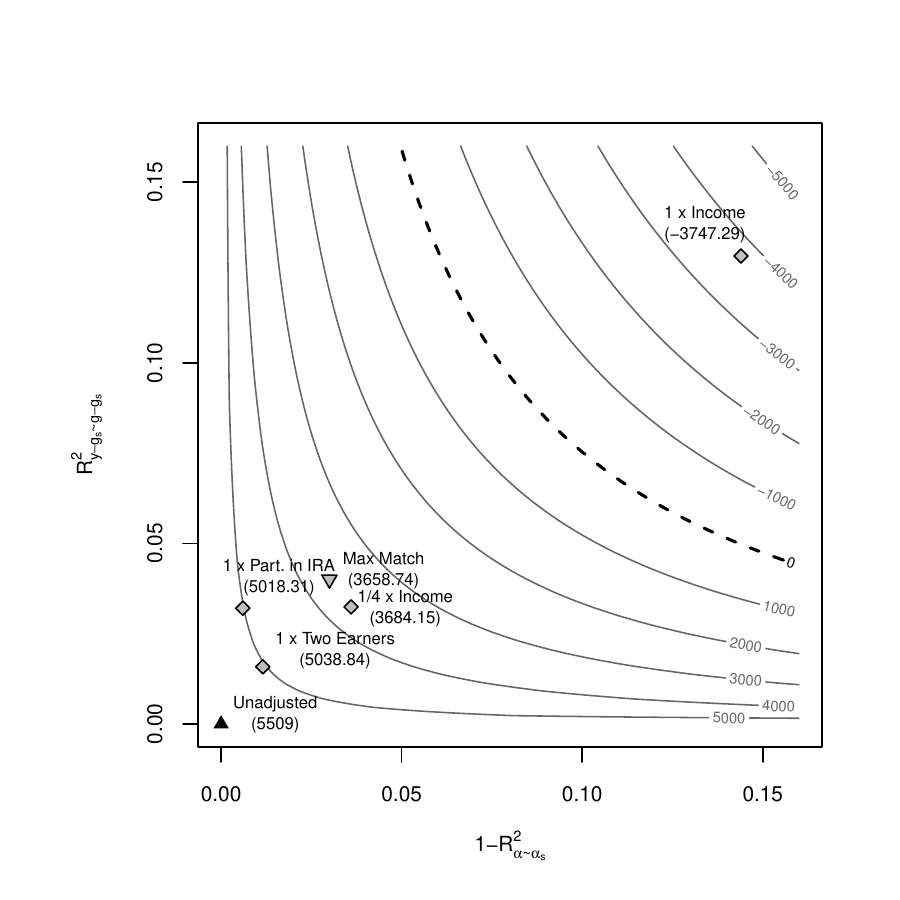}
    \caption{\small Lower limit confidence bound, NPM, $|\rho|=1/2$.}
    \label{fig:npm-lower-rho}
\end{subfigure}
\caption{\small Sensitivity contour plots 401(k). Significance level $a=0.05$.} 
\label{fig:contours-plm-bounds}
\end{figure}

Starting with the partially linear model, the results are shown in Figure~\ref{fig:plm-lower}. The horizontal axis describes the fraction of residual variation of the treatment explained by unobserved confounders, whereas the vertical axis describes the share of residual variation of the outcome explained by unobserved confounders. The contour lines show the lower limit of the confidence bounds $[l, u]$ for the ATE (see Theorem~\ref{thm:DML}), given a pair of hypothesized values of partial $R^2$. Note $\text{RV}_{\theta=0, a=0.05}$ of Table~\ref{tab:minimal} is simply the point where the 45-degree line crosses the contour line of zero (red dashed line), offering a convenient summary of the critical contour. We can further place  reference points on the contour plots, indicating plausible bounds on the strength of confounding, under alternative assumptions about the maximum explanatory power of omitted variables. The grey triangle point on the plot---\emph{Max Match}---shows the bounds on the partial $R^2$ as previously discussed, resulting in a lower limit of the confidence bound for the ATE of \$2,497, in accordance with Table~\ref{tab:401k}. Note here the correlation $|\rho|$ is set to its upper bound of 1. 

Another approach to construct confounding scenarios is to use observed covariates to bound the plausible strength of unobserved covariates. For instance, in our empirical example,  we know that employment decisions are largely driven by salary considerations. Similarly, salary is clearly an important determinant of net financial assets. One could therefore argue that it is implausible to imagine other latent firm characteristics that would be even a fraction as  strong as  the observed \emph{income} of individuals, in terms of explanatory power in predicting 401(k) eligibility and net financial assets. Whenever such claims of relative importance can be made, they can be used to set plausible bounds on the strength of unmeasured confounding. Formal details of this benchmarking procedure are provided in the Online Appendix.\footnote{Since the benchmarks are estimated from data, one could also propagate the uncertainty due to estimation errors of the benchmarks via the delta method.}

The grey diamonds of Figure~\ref{fig:plm-lower} shows the bounds on the strength of the latent variable $U$ if it were as strong as  (i)~income (\emph{1 x Income}), (ii) whether a worker has an individual retirement account \mbox{(\emph{1 x Part. in IRA})}, and (iii) whether the worker's family has a two-earner status (\emph{1 x Two Earners}). Note that, apart from income, latent variables as strong as these covariates would result in a weaker confounding scenario than the one we have previously considered (\emph{Max Match}).  As for income, the worst-case bound indicates that omitted firm characteristics as important as income could be sufficiently strong to overturn the original results. However, one could argue such scenario to be implausible, as it is hard to imagine latent firm characteristics that would explain more variation in job choice than income itself. A more realistic, but still conservative, scenario is thus provided by the benchmark point \emph{1/4 x Income}, which shows the bound on the strength of $U$ if it were 25\% as strong as income in predicting treatment and outcome variation. Note this scenario is comparable to the \emph{Max Match} scenario, and not enough to bring the lower limit of the confidence bound to zero.


All results of Figure~\ref{fig:plm-lower} were computed under the very conservative assumption that, given a pair of partial $R^2$ values for the latent variable $U$, the confounders enter both the outcome and treatment equations in a way that maximizes the bias, resulting in $|\rho|=1$. This may be a conservative scenario, especially in nonlinear models. Thus, similar benchmarking procedures used for assessing the plausibility of the $R^2$ values can also be employed to calibrate judgments on the magnitude of $\rho$. In fact none of the observed covariates result in $|\rho|$ values~exceeding~1/2. With this in mind, Figure~\ref{fig:plm-lower-rho} presents the same contour plots as before, but now with $|\rho|$ set to a less conservative value of 1/2. Note how this substantially attenuates the bias,  with the lower limits of the confidence bounds reaching approximately $\$4,600$ and $\$5,400$ for the \emph{Max Match} and \emph{1/4 x Income}, respectively.  

Sensitivity contour plots for the nonparametric model are similar but slightly more conservative, and are provided in Figures~\ref{fig:npm-lower}~and~\ref{fig:npm-lower-rho}. The interpretation of the contours is the same as before, with the main difference being that the horizontal axis now describes gains in precision instead of gains in variance explained (see, e.g., Remark~\ref{rmk:ate}).

\section{Conclusion}\label{sec:conclusion}

In this paper we provide a general theory of omitted variable bias for continuous linear functionals of the conditional expectation function of the outcome---all for general, nonparametric, causal models, while naturally allowing for (semi-)parametric restrictions (such as partial linearity), when such assumptions are made. We allow for arbitrary (e.g., binary or continuous) treatment and outcome variables, and we show that the bounds on the bias depend only on the  maximum explanatory power of latent variables. We provide theoretical details of many  leading causal estimands, and, in particular, we derive novel results for the important special cases of average treatment effects in partially linear models, in nonparametric models with a binary treatment, as well as for average causal derivatives.  Finally, we leverage the Riesz representation of our bounds to offer flexible statistical inference through (debiased) machine learning, with rigorous coverage guarantees.

Many interesting extensions of our work are possible. For example, our results can potentially be extended to nonlinear functionals, such as those arising in instrumental variable (IV) methods. For instance, consider a variant of the IV problem \citep{imbens:angrist:94}, where the instrumental variable $Z$ is valid only when conditioning both on observed covariates $X$, and latent variables $U$. In this case, the IV estimand is given by the ratio of two average treatment effects,  and both the numerator and denominator can be bounded using the methods for the ATE proposed in this paper.  Another interesting direction is to consider causal estimands that are functionals of the long quantile regression, or causal estimands that are values of a policy in dynamic stochastic programming. When the degree of confounding is small, it seems possible to use the results in \cite{DML-LR} to derive approximate bounds on the bias that can be estimated using debiased ML approaches.  Finally, our results  can potentially be extended to the richer class of ``mixed bias'' functionals \citep{rotnitzky2021characterization}, which includes ``doubly robust'' functionals \citep{robins2008higher,cui2024selective} as a special case.

\section*{Acknowledgments, Data Availability, Conflict of Interests, and Funding}

\paragraph{\textbf{Acknowledgments}} 
We thank the Editor and anonymous reviewers for very helpful comments. We are also grateful to Isaiah Andrews, Elias Bareinboim, Ben Deaner, Jacob Dorn, Xavier Gabaix, David Green, Guido Imbens, Steve Lehrer, Judith Lok, Esfandiar Maasoumi, Anna Mikusheva, Richard Nickl, Judea Pearl, Jack Porter, James Poterba, Alejandro Schuler, Eric Tchetgen Tchetgen, Ingrid Van Keilegom, and the participants of the Chamberlain seminar, Online Causal Inference Seminar, Canadian Economic Association meeting, Institute for Nonparametric Statistics meeting, Bernoulli–IMS World Congress, EuroCIM, ACIC, PCIC, SBE, JSM, PolMeth, and UAI, as well as participants of seminars at Harvard-MIT, Stanford, Wisconsin, Emory, Berkeley Methods Workshop, University of Washington, Cornell, Syracuse University, Penn State University, and BU Causal Seminar for their valuable feedback. We are grateful to Jack Porter for suggesting the ``long story short'' title. As usual, all remaining errors are our own.

\paragraph{\textbf{Data availability.}} All data is publicly available in our GitHub repository: \url{https://github.com/carloscinelli/dml.sensemakr}. An R package for the methods developed in this paper is also available at the same repository. Alternative R and Python implementations are  available via the DoubleML ecosystem at  \url{https://docs.doubleml.org/stable/index.html}.

\paragraph{\textbf{Conflict of interest.}} There are no relevant financial or nonfinancial competing interests to report.

\paragraph{\textbf{Funding.}} This research was partially funded by the Royalty Research Fund at the University of Washington and by the National Science Foundation under Grants  No. MMS-2417955 and  2242447.

\clearpage
\setcounter{page}{1}

\appendix

\section*{\textbf{Online Appendix for\\ ``Long Story Short: Omitted Variable Bias in Causal Machine Learning''}}

\section{Theoretical details for leading causal estimands}\label{sec:leading-examples}

In this section we provide theoretical details for a wide variety of interesting and important causal estimands. Recall that we use $W = (D,X,U)$ to denote the ``long'' set of regressors and  $W_s = (D,X)$ to denote the ``short'' list of regressors. For all estimands below, we consider the fully nonparametric regressions $g(D, X,U):=\Ep[Y\mid D,X,U]$ and $g_s(D,X):= \Ep[Y\mid D,X]$.

Let us start with examples for the binary treatment case, with the understanding that finitely discrete treatments can be analyzed similarly.  

\begin{example}[\textbf{Weighted Average Potential Outcome}]    
\label{ex:wpo}
Let $D \in \{0,1\}$ be the indicator of the receipt of the treatment. Define the long parameter as
$
\theta = \Ep[ g(\bar d, X,U) \ell(W_s)],$
where $W_s  \mapsto \ell(W_s)$ is a bounded non-negative weighting function and $\bar d$ is a fixed value in $\{0,1\}$. We define the short parameter as 
$
\theta_s = \Ep[g_s(\bar d, X) \ell(W_s)].$
We assume  $\Ep Y^2< \infty$ and the weak overlap condition $ \Ep[\ell^2(W_s)/P(D= \bar d \mid X,U)] < \infty.$
\end{example} 

The long parameter is a weighted average potential outcome (PO) when we set the treatment to $\bar d$, under the standard conditional ignorability assumption (\ref{eq: CE}). The short parameter is a statistical approximation based on the short regression. In this example, setting \begin{itemize}
\item $\ell(W_s)  = 1$ gives the average PO in the entire population; 
\item $\ell(W_s) = 1( X \in {\mathcal{N}})/P(X \in \mathcal{N})$  the average PO for group $\mathcal{N}$;
\item $\ell(W_s)= D/P(D=1)$ the average PO for the treated;
\item $\ell(W_s)= (1-D)/P(D=0)$ the average PO for the untreated.
\end{itemize}
Above we can consider $\mathcal{N}$ as small regions shrinking in volume with the sample size, to make the averages local, as in \cite{RR-local}, but for simplicity we take them as fixed in this paper. 

\begin{example}[\textbf{Weighted Average Treatment Effects}]
\label{ex:att}
  In the setting of the previous example, define the long parameter $
\theta = \Ep [(g(1,X,U) - g(0,X,U))  \ell(W_s)],$
and the short parameter as 
$
\theta_s = \Ep[(g_s(1,X) - g_s(0,X)) \ell(W_s)].$ 
We further assume $\Ep Y^2< \infty$ and the weak overlap condition $\Ep[\ell^2(W_s)/\{P(D= 0 \mid X,U) P(D= 1 \mid X,U)\}] < \infty. $ \end{example} 

The long parameter is a weighted average treatment effect under the standard conditional ignorability assumption. In this example, setting 
\begin{itemize}
\item $\ell(W_s)  = 1$ gives ATE in the entire population; 
\item $\ell(W_s) = 1( X \in \mathcal{N})/P(X \in \mathcal{N})$ the ATE for group $\mathcal{N}$;
\item $\ell(W_s) = D/P(D=1)$ the ATE for the treated (ATT);
\item $\ell(W_s) = (1-D)/P(D=0)$ the ATE for the untreated (ATU);
\item $\ell(x) = \pi(x)$ 
 the average value of policy (APV) $\pi$, 
\end{itemize}
where the policy $\pi$ assigns a fraction $0\leq \pi(x) \leq 1$ of the subpopulation with observed covariate value $x$ to receive the treatment. 

In what follows $D$ does not need to be binary. We next consider a weighted average  effect of changing observed covariates $W_s$ according to  a transport map $W_s \mapsto T(W_s)$, where $T$ is deterministic measurable map from $\mathcal{W}^s$ to $\mathcal{W}^s$.  For example, the policy  $$
(D,X,U) \mapsto (D +  1,  X,U)$$ adds a unit to the treatment $D$, that is $T(W_s) = (D+1,X)$. This has a causal interpretation if the policy induces the equivariant change in the regression function, namely  the counterfactual outcome $\tilde Y$ under the policy obeys $\Ep [\tilde Y \mid X,U] = g(T(W_s),U)$, and the counterfactual covariates are given by $\tilde W = (T(W_s),U)$.

\begin{example}[\textbf{Average Policy Effect from Transporting $W_s$}]
\label{ex:policy} 
For a bounded  weighting function $W_s \mapsto \ell(W_s)$, the long parameter is given by
$
\theta = \Ep[ \{g(T(W_s),U) - g(W_s,U)\} \ell(W_s) ].$ 
The short form of this parameter is 
$
\theta_s = \Ep[ \{g_s(T(W_s)) - g_s(W_s)\} \ell(W_s) ]. $
As the regularity conditions we require that the support of  $P_{\tilde W} = \mathrm{Law}(T(W_s),U)$ is included in the support of $P_W$, and require the weak overlap condition $
\Ep[\alpha(W)^2]< \infty$, where $\alpha$ is the long Riesz representer given in Theorem~\ref{thm:ovb-validity}.
 \end{example} 

We now turn to examples with continuous treatments $D$ taking values in $\mathbb{R}^k$. 
Consider the  average causal effect  of the policy that shifts the distribution of covariates via the map $W=(D,X,U) \mapsto (T(W_s),U) = (D +  r t(W_s), X,U )$  weighted by $\ell(W_s)$, keeping the long regression function invariant.  The following long parameter $\theta$ is an approximation to  $1/r$ times  this average causal effect for small values of $r$. This example is a differential version of the previous example. 

\begin{example}[\textbf{Weighted Average Incremental Effects}] 
\label{ex:waie}
Consider the long parameter taking the form of the average directional derivative:
$
\theta =  \Ep [\ell(W_s) t(W_s) '\partial_d g(D,X,U)],$
where $\ell$ is a bounded weighting function and $t$ is a bounded direction function. The short form of this parameter is
$
\theta_s =  \Ep [\ell(W_s) t(W_s) '\partial_d g_s(D,X)].
$
As regularity conditions, we suppose that $\Ep Y^2< \infty$. Further for each $(x,u)$ in the support of $(X,U)$, and each $d$ in $ \mathcal{D}_{x,u}$, the support of $D$ given $(X,U) = (x,u)$, the derivative maps $d \mapsto \partial_d g(d,x,u)$ and $d \mapsto g(w) \omega(w)$,  for $\omega(w) := \ell(d,x) t(d,x) f(d\mid x,u)$, are continuously differentiable; the set $\mathcal{D}_{x,u}$ is bounded, and its boundary is piecewise-smooth; and $\omega(w)$ vanishes for each $d$ in this boundary. Moreover, we assume the weak overlap: $ \Ep [(\divg_d  \omega(W) /f(D\mid X,U))^2] < \infty.$
We also assume that the expectations defining $\theta$ and $\theta_s$ are finite.
\end{example}

Another example is that of a policy that shifts the entire distribution of observed covariates, independently of $U$.  The following long parameter corresponds to the average causal contrast  of two  policies that set the distribution of observed covariates $W_s$ to $F_0$ and $F_1$,  independently of $U$.  Note that this example is different from the transport example, since here the dependence between $U$ and $W_s$ is eliminated under the interventions.

\begin{example}[\textbf{Policy Effect from Changing Distribution of $W_s$}]
\label{ex:pecd}
Define the long parameter as $
\theta =  \int \left [\int g(W_s,u)d P_U(u) \right] \ell(W_s)   d \mu(W_s); \quad \mu(W_s) =  F_1(W_s) - F_0(W_s),
$
where $\ell$ is  a bounded weight function, and the short parameter as
$
\theta_s =  \int g_s(W_s) \ell(W_s)   d \mu(W_s); \quad \mu(W_s) =  F_1(W_s) - F_0(W_s).
$
As the regularity conditions we require that the supports of $F_0$ and $F_1$ are contained in the support of $W_s$, and that the measure $d P_U \times d F_k$ is absolutely continuous with respect to the measure $d P_W$ on $\mathcal{U} \times \text{support} (\ell)$. We further assume that $\Ep Y^2< \infty$ and the weak overlap: $
\Ep [(\ell [d P_U \times d (F_1 - F_0)]/d P)^2]< \infty. $
\end{example}

The following result establishes the validity of the OVB formula for all examples.

\begin{Theorem}[\textbf{OVB Validity in Examples 1-5}]\label{thm:ovb-validity} Under the conditions stated in Examples 1,2,3,5, Assumptions 1 and 2 are satisfied.  Under conditions stated in Example 4, Assumptions 1 and 2 are satisfied for the Hahn-Banach extension of the mapping $g \mapsto \Ep m(W,g)$ to the entire $L^2(P_W)$, given by $g \mapsto \Ep g(W) \alpha(W)$. The scores for Examples 1-5 are given by:

\hspace{-1em}{\small
\begingroup
\setlength{\tabcolsep}{-1ex} 
\begin{tabular}{p{.55\textwidth}p{.55\textwidth}}
    \begin{enumerate}
\item  $m(w, g)  = (g(\bar d,x,u))\ell(W_s)$;  \item  $m(w, g)  = (g(1,x,u) - g(0,x,u)) \ell(W_s) $;  
\item  $m(w,g) = (g(T(W_s),u) - g(W_s,u)) \ell(W_s)$;
\item  $m(w, g) =  \ell(W_s) t(W_s)'\partial_d g(w)$;
\item $m (w, g) = \int [\int g(W_s,u) dP_U(u)] \ell(W_s) d \mu(W_s)$;
 \end{enumerate}
     & 
     \begin{enumerate}
\item  $m(W_s, g_s)  = (g_s(\bar d,x))\ell(W_s)$;  \item  $m(W_s, g_s)  = (g_s(1,x) - g_s(0,x)) \ell(W_s) $;  
\item  $m(w_s,g) = (g_s(T(W_s)) - g_s(W_s)) \ell(W_s)$;
\item  $m(W_s, g_s) =  \ell(W_s) t(W_s)'\partial_d g_s(W_s)$;
\item $m (W_s, g_s) = \int g_s(W_s) \ell(W_s) d \mu(W_s)$.
 \end{enumerate}
\end{tabular}
\endgroup}
The long RR and corresponding short RR are given by:

\hspace{-1em}{\small
\begingroup
\setlength{\tabcolsep}{-1ex} 
\begin{tabular}{p{.6\textwidth}p{.6\textwidth}}
    \begin{enumerate}
\item $\alpha(w)  =    \ \frac{1(d=\bar d)}{p(\bar d\mid x,u)} \bar \ell(x,u);$
\item $\alpha(w) = \ \frac{1(d=1) - 1(d= 0)}{p(d\mid x,u)} \bar \ell(x,u);$
\item $\alpha(w)  =  \frac{d P_{\tilde W}(w)\Ep[\ell(W_s)|\tilde W]- dP_{W}(w)\ell(W_s)}{d P(w)}$;
\item $\alpha(w) = - \frac{\divg_d ( \ell(W_s) t(W_s) f(d\mid x,u))}{f(d\mid x,u)}$;
\item $ \alpha(w)  =    \ \frac{d P_U(u) \times d (F_1(W_s) - F_0(W_s))}{d P(w)}  \ell(W_s); \label{rr: ex2}$ 
\end{enumerate}
     & 
     \begin{enumerate}
\item $\alpha_s(W_s)  =    \ \frac{1(d=\bar d)}{p(\bar d\mid x)} \bar \ell(x);$
\item $\alpha_s(W_s) = \ \frac{1(d=1) - 1(d= 0)}{p(d\mid x)} \bar \ell(x);$
\item $\alpha_s(W_s)  =  \frac{d P_{\tilde W_s}(W_s)\Ep[\ell(W_s)|\tilde W_s] - dP_{W_s}(W_s) \ell(W_s)}{d P_{W_s}(W_s)} $
\item $\alpha_s(W_s) = - \frac{\divg_d ( \ell(W_s) t(W_s) f(d\mid x) )}{f(d\mid x)}$;
\item $ \alpha_s(W_s) =   \ \frac{d (F_1(W_s) - F_0(W_s))}{d P_{W_s}(W_s)}  \ell(W_s); \label{rrs: ex2}$ 
\end{enumerate}
\end{tabular}
\endgroup}
where above we used the notations: $\bar \ell (X,U) := \Ep [\ell(W_s)\mid X,U], \bar \ell (X) := \Ep [\ell(W_s)\mid X]$, $ p(d\mid x,u) := \Pr (D=d\mid X=x, U=u), p(d\mid x) := \Pr (D=d\mid X=x).$ In Examples 1-2, when the weight function only depends on $X$,  namely $\ell(W_s) = \ell(X)$,
we have the simplifications $\bar \ell (X,U) = \bar \ell(X) = \ell(X).$
\end{Theorem}

As we have seen in Remarks~\ref{rmk:ate},~\ref{rmk:att},~and~\ref{rmk:acd}, it may be useful to further specialize the interpretation of the sensitivity parameters  $1-R^2_{\alpha\sim\alpha_s}$ for the many cases encompassed by the examples of Theorem~\ref{thm:ovb-validity}. As this would be an extensive task, we leave such specializations to future work.

\subsection*{Statistical Inference.}

The examples above involve an extra nuisance function, such as the weighting function $\ell(W_s)$. If $\ell(W_s)$ is treated as a known function, then statistical inference can be conducted as before as in Theorem~\ref{thm:DML}. If  $\ell(W_s)$ is estimated from data, then its sampling variability needs to also be taken into account.  For instance, in the case of the ATE, ATT and ATU of Example~\ref{ex:att}, the influence function for $\theta_s$ becomes,
\begin{equation}\label{eqn:att-influence}
\psi_{\theta_s} (Z; \theta_s, g_s, \alpha_s, \ell) := m(W_s, g_s) + (Y- g_s(W_s)) \alpha_s(W_s) - \theta_s \times \ell(W_s).
\end{equation}

More generally, if we have a moment of the form:
$$
m(W_s,g_s): = m^*(W_s,g_s)\times \ell(W_s),
$$
where the score $m^*(W_s, g_s)$ does not depend on any unknown nuisance function, and the weighting function $\ell(W_s)$ is known up to some unknown nuisance component $\eta_\ell(V)$, with $V$ being a subset of the variables $W_s$, i.e. $\ell(W_s, \eta_\ell(V))$, then the influence function for $\theta_s$ also needs to take into account the influence from estimating $\eta_\ell$. If $\eta_\ell$ corresponds to the solution to some other regression problem, i.e. $\Ep[Q\mid V] = \eta_\ell(V)$, with $Q$ being an observed variable, then the influence function for $\theta_s$ becomes,
$$
\psi_{\theta_s} (Z; \theta_s, g_s, \alpha_s, \ell) := m(W_s, g_s) + (Y- g_s(W_s)) \alpha_s(W_s) + (Q - \eta_\ell(V))\,\mu(V) - \theta_s.
$$
where $\mu(V) := \Ep[m^*(W_s,g_s)\,\partial_{\eta_\ell}\ell(W_s, \eta_\ell(V))\mid V]$. The ATT is a special case of this construction 
where $\eta_\ell=P(D=1)$, 
$V$ is the empty set of variables, 
$\ell(W_s, \eta_\ell)=D/P(D=1)$, 
$Q=D$ and 
$\mu(V) = -\Ep[m^*(W_s,g_s) D / P(D=1)^2] = -\theta_s / P(D=1)$. 
Hence, the extra correction term takes the form $-\theta_s (D - P(D=1))/P(D=1)$, which can be combined with the term $\theta_s$ to yield the term $-\theta_s D/P(D=1)=-\theta_s \times \ell(W_s)$ that we provide in Equation~\eqref{eqn:att-influence}.


\section{Preliminaries}

\subsection{Few Preliminaries}
\label{app:prelim}
To prove supporting lemmas we recall the following standard definitions and results. Given two normed vector spaces $V$ and $W$ over the field of  real numbers $\mathbb{R}$, a linear map $A : V \to W$ is continuous if and only if it has a bounded operator norm:
$$\displaystyle \|A\|_{op} :=\inf\{c\geq 0:\|Av\|\leq c\|v\|{\mbox{ for all }}v\in V\} < \infty,$$
where $\| \cdot \|_{op}$ is the operator norm. The operator norm depends on the choice of norms for the normed vector spaces $V$ and $W$.  A Hilbert space is a complete linear space equipped with an inner product $\langle f ,  g \rangle$
and the norm $ |\langle f,  f \rangle|^{1/2}$. The space $L^2(P)$ is the Hilbert space with the inner product
$\langle f, g \rangle = \int f g dP$ and norm $\|f\|_{P,2}$. The closed linear subspaces of $L^2(P)$ equipped with the same inner product and norm are Hilbert spaces.

\textbf{Hahn-Banach Extension for Normed Vector Spaces.}  If $V$ is a normed vector space with linear subspace $U$ (not necessarily closed) and if $\phi : U \mapsto K$ is continuous and linear, then there exists an extension $\psi: V \mapsto K$ of $\phi$ which is also continuous and linear and which has the same operator norm as $ \phi$.

\textbf{Riesz-Frechet Representation Theorem.} Let $H$ be a Hilbert space over $\mathbb{R}$ with an inner product $\langle \cdot , \cdot \rangle$, and $T$ a bounded linear functional mapping $H$ to $\mathbb{R}$. If $T$ is bounded then
there exists a unique $g \in H$ such that for every $f \in H$ we have $T(f) =\langle f, g\rangle$. It is given
by $g = z (T z)$, where $z$ is unit-norm element of the orthogonal complement of the kernel subspace
$K = \{ a \in H:  T a = 0\}$. Moreover, $\|T\|_{op} = \|g\|$,
where $\|T\|_{op}$ denotes the operator norm of $T$, while $\|g\|$ denotes the Hilbert space norm of $g$.

\textbf{Radon-Nikodym Derivative.}  Consider a measure space $(\mathcal{X},\mathcal{A})$ 
on which two $\sigma$-finite measures  are defined, $\mu$ and $\nu$.
If $\nu \ll \mu$ (i.e. $\nu$ is absolutely continuous with respect to $\mu$), then there is a  measurable function $f: \mathcal{X} \rightarrow [0,\infty) $, such that for any measurable set $A \subseteq \mathcal{X} $,
$\nu(A) = \int_A f \, d\mu$. The function $f$ is conventionally denoted by $d\nu/d\mu$.

\textbf{Integration by Parts.} Consider a closed measurable subset $\mathcal{X}$ of $\mathbb{R}^k$ equipped with Lebesgue measure $V$ and piecewise smooth boundary $\partial \mathcal{X}$, and suppose that $v: \mathcal{X} \to \mathbb{R}^k$ and 
$\phi: \mathcal{X} \to \mathbb{R}$ are both $C^1(\mathcal{X})$,
then 
$$
 {\displaystyle \int _{\mathcal{X} }\varphi \operatorname {div} { {v}}\,dV=\int _{\partial \mathcal{X} }\varphi \,{ {v}}'n d{ {S}}-\int _{\mathcal{X} }{ {v}}' \operatorname {grad} \varphi \,dV,}
$$
where $S$ is the surface measure over the surface $\partial \mathcal{X}$ induced by $V$, and $n$ is the outward normal vector.

\section{Deferred Proofs and Examples}

\subsection{Proof of Theorem \ref{thm:ovb-plm}}
The result follows from 
\begin{eqnarray*}
\Ep g \alpha - \Ep g_s \alpha_s
 &= &  \Ep (g_s + g- g_s) (\alpha_s + \alpha - 
\alpha_s) - \Ep g_s \alpha_s \\
& = & \Ep g_s (\alpha- \alpha_s) +
\Ep \alpha_s (g - g_s) + \Ep (g-g_s) (\alpha-\alpha_s) \\
& =&  \Ep (g-g_s) (\alpha-\alpha_s),
\end{eqnarray*}
using the fact that $\alpha_s$ is orthogonal to $g- g_s$
and $g_s$ is orthogonal to $\alpha-\alpha_s$.

Moreover, 
$$
\Ep(g - g_s)(\alpha - \alpha_s) = \frac{\Ep(g - g_s)(\alpha - \alpha_s) }{\sqrt{\Ep(g - g_s)^2\Ep(\alpha - \alpha_s)^2} } \times \sqrt{\Ep(g - g_s)^2 \Ep(\alpha - \alpha_s)^2}  = \rho B,
$$
where
$$
\rho:= \frac{\Ep(g - g_s)(\alpha - \alpha_s) }{\sqrt{\Ep(g - g_s)^2\Ep(\alpha - \alpha_s)^2} }, 
\quad
B := {\sqrt{\Ep(g - g_s)^2\Ep(\alpha - \alpha_s)^2} }.
$$
Since $\Ep(g-g_s) = \Ep(\alpha-\alpha_s)=0$, we also have $\rho = \mathrm{Cor}(g-g_s, \alpha-\alpha_s)$.

We further note that $B^2$ factorizes as 
$
B^2 =  C^2_Y C^2_DS^2,
$
where $S^2  :=  \Ep (Y-g_s)^2 \Ep \alpha_s^2$, and 
$$
C^2_Y = \frac{\Ep( g- g_s)^2}{\Ep (Y- g_s)^2} = R^2_{Y-g_s \sim g- g_s},
$$
and 
$$
C^2_D =
\frac{\Ep( \alpha- \alpha_s)^2}{\Ep \alpha_s^2} =
\frac{\Ep \alpha^2-  \Ep \alpha^2_s}{\Ep \alpha_s^2}
=\frac{ 1/\Ep \tilde D^2 - 1/\Ep \tilde D_s^2}{ 1/\Ep \tilde D_s^2} = 
\frac{\Ep \tilde D^2_s - \Ep \tilde D^2 }{\Ep \tilde D^2} = \frac{R^2_{\tilde D_s \sim \tilde U }}{1 - R^2_{\tilde D_s \sim \tilde U} },
$$
where $\tilde D := D - \Ep[D \mid X, U]$,
$\tilde D_s := D - \Ep[D \mid X]$, and $\tilde U = \Ep[D \mid X, U]- \Ep[D \mid X]$. 

Here  we used the observation that
$$
\Ep (\alpha - \alpha_s)^2= \Ep \alpha^2 + \Ep \alpha^2_s - 2 \Ep \alpha \alpha_s =\Ep \alpha^2 -  \Ep \alpha^2_s,
$$
holds because
$$
\Ep \alpha \alpha_s = \frac{ \Ep \tilde D \tilde D_s }{\Ep \tilde D^2 \Ep \tilde D_s^2} = \frac{\Ep \tilde D^2}{\Ep \tilde D^2 \Ep \tilde D_s^2} = \frac{1}{\Ep \tilde{D}^2_s} = \Ep \alpha^2_s.
$$
In addition, we note 
$$
\frac{\Ep \alpha^2-  \Ep \alpha^2_s}{\Ep \alpha_s^2}
= \frac{\Ep \alpha^2-  \Ep \alpha^2_s}{\Ep \alpha^2} \frac{\Ep \alpha^2}{\Ep \alpha_s^2} = \frac{1- R^2_{\alpha \sim \alpha_s}}{ R^2_{\alpha \sim \alpha_s}}, 
$$
and,
$$
R^2_{\tilde D_s \sim \tilde U} = \eta^2_{D \sim U\mid X}. 
$$

Finally, under correct specification,  
$$
\Ep(Y-g_s)^2 = \Ep[\Var(Y\mid D, X)], \quad R^2_{Y -g_s \sim g-g_s} = \eta^2_{Y\sim U\mid D, X}. \qed
$$

\subsection{Relationship to the linear OVB formula}
\label{sec:app-linear-ovb}
The result of Theorem~\ref{thm:ovb-plm} generalizes the traditional OVB formula for linear models. To see how, without loss of generality, consider centered random variables, and scalar $X$ and $U$. Now let,
$$
g := \theta D + \beta X + \gamma U, \qquad g_s:=\theta_sD + \beta_s X.
$$
Here, as before, the models need not be correctly specified, and $g, g_s$ can simply be the best linear approximations of the long and short CEF. Note the regression error is given by,
$$
g - g_s = (\theta -\theta_s)D + (\beta-\beta_s)X + \gamma U,
$$
and the error in the Riesz representer is given by,
$$
\alpha -\alpha_s = \frac{\tilde D}{\mathrm{Var}(\tilde D)} - \frac{\tilde D_s }{\mathrm{Var}(\tilde D_s)},
$$
where here we write $\tilde D$ to denote $D$ after partialling out $X$ and $U$, and  $\tilde D_s $ to denote $D$ after partialling out $X$ only. Now note $\Ep[D\tilde D] = \mathrm{Var}(\tilde D)$  and $\Ep[D \tilde D_s] = \mathrm{Var}(\tilde D_s)$. Moreover, $\Ep[\tilde DX]= \Ep[\tilde D_sX]=0$.

Thus, when taking the covariance of the two errors, all that is left out is the term containing the  covariance of $U$ with $\tilde D_s$:
\begin{align*}
\theta - \theta_s = \Ep[(g - g_s)(\alpha - \alpha_s)] = - \gamma \times \underbrace{\frac{\Ep[U\tilde D_s]}{\mathrm{Var}(\tilde D_s)}}_{\delta}= - \gamma \delta.  
\end{align*}
By the FWL theorem, $\delta$ is nothing but the regression coefficient of $D$ on the regression of $U$ on $D$, adjusting for $X$, thus,
$$
\theta - \theta_s = - \gamma \delta
$$
which equals the traditional omitted variable bias formula for linear models presented in textbooks \citep{goldberger1991course,AngristBook,wooldridge:text}.

\subsection{Example of sharpness for PLM}
\label{app:sharp-plm}

Suppose the observed data distribution, $P(Y,D,X)$ factorizes as,
\begin{align*}
X &\sim P(X),\\
D\mid X &\sim \mathcal{N}(\mu_d(X),~ \sigma^2_d ),\\
Y\mid D, X &\sim \mathcal{N}(\theta_s\left(D-\mu_d(X)\right) + \mu_y(X),~ \sigma^2_y ),
\end{align*}
where $\sigma^2_y > 0$, $\sigma^2_d >0$, and $\mathcal{N}(\mu,\sigma^2)$ denotes the density of a standard normal with mean $\mu$ and variance $\sigma^2$. 

We will show that we can always construct a full data law $P(Y, D, X, U)$ such that: (a) the long regression model is also partially linear on $D$; (b) the full data law marginalizes to the observed data law $P(Y, D, X)$; and (c) we achieve any value  $\rho \in [-1,1]$, $\eta^2_{D\sim U\mid X} <1$, $\eta^2_{Y\sim U\mid D,X} < 1$.

Our proof is constructive. Let the full data law $P(Y, D, X, U)$ be:
\begin{align*}
U_1 &\sim \mathcal{N}(0,1)\\
U_2\mid U_1 &\sim \mathcal{N}(0,1)\\
X\mid U_2,U_1 &\sim P(X),\\
D\mid X, U_1, U_2 &\sim \mathcal{N}(\mu_d(X) + \delta U_1,~ (1-\eta^2_d)\times \sigma^2_d ),\\
Y\mid D, X, U_1, U_2 &\sim \mathcal{N}(\theta\left(D-\mu_d(X)\right) + \mu_y(x) + \gamma_1 U_1 + \gamma_2 U_2,~  (1-\eta^2_y)\times \sigma^2_y),
\end{align*}
where, 
$$
\theta = \theta_s + \rho\sqrt{\frac{\eta^2_y\eta^2_d}{1-\eta^2_d}  \frac{\sigma^2_y}{\sigma^2_d}} , 
\quad \delta = \sqrt{\eta^2_d \sigma^2_d}, 
\quad \gamma_1 = -\rho\sqrt{\frac{\eta^2_y \sigma^2_y}{1-\eta^2_d}}, 
\quad \gamma_2 = \sqrt{(1-\rho^2) \eta^2_{y} \sigma^2_y}.
$$
We define $U = (U_1, U_2).$ We first show that this distributions marginalizes to the observed data. First, since $U_1$ and $U_2$ are jointly independent Gaussian, this implies $D\mid X$ and $Y\mid D,X$ are also Gaussian. It thus suffices to show that the marginal means and variances are preserved. 

Starting with $D\mid X$:
\begin{align*}
\Ep[D\mid X] &= \Ep[\mu_d(X) + \delta U_1\mid X] = \mu_d(X) + \delta \Ep[U_1\mid X] = \mu_d(X),\\    
\Var(D\mid X) &= \delta^2 \Var(U_1\mid X) + (1-\eta^2_d)\sigma_d^2 = \eta_d^2\sigma^2_d + (1-\eta^2_d)\sigma_d^2 = \sigma^2_d.
\end{align*}
That is the marginal distribution of $D\mid X$ matches that of the observed data. Now moving to $Y\mid D,X$, first note that:
$$
U_1\mid D,X \sim \mathcal{N}\left(\sqrt{\frac{\eta^2_d}{\sigma^2_d}}(D-\mu_d(X)),~ 1-\eta_d^2\right)
$$
and $U_2 \indep \{U_1, D, X\}$. Thus, for the CEF we obtain,
\begin{align*}
\Ep[Y\mid D, X] &= \Ep[\theta\left(D-\mu_d(X)\right) + \mu_y(x) + \gamma_1 U_1 + \gamma_2 U_2\mid D, X] \\
&= \theta\left(D-\mu_d(X)\right) + \mu_y(x) + \gamma_1\Ep[U_1\mid D, X] + \gamma_2 \Ep[U_2\mid D,X]\\
&= \theta\left(D-\mu_d(X)\right) + \mu_y(x) + \gamma_1 \times \sqrt{\frac{\eta^2_d}{\sigma^2_d}}(D-\mu_d(X))\\
&= \left(\theta_s + \rho\sqrt{\frac{\eta^2_y\eta^2_d}{1-\eta^2_d} \frac{\sigma^2_y}{\sigma^2_d}}\right)\left(D-\mu_d(X)\right) + \mu_y(x) \\
&\qquad \quad -\rho\sqrt{\frac{\eta^2_y \sigma^2_y}{1-\eta^2_d}} \times \sqrt{\frac{\eta^2_d}{\sigma^2_d}}(D-\mu_d(X))\\
&= \theta_s \left(D-\mu_d(X)\right) + \mu_y(x).
\end{align*}
And for the variance,
\begin{align*}
    \Var(Y\mid D,X) &= \gamma_1^2 \Var(U_1\mid D,X) + \gamma_2^2\Var(U_2\mid D,X) + (1-\eta^2_y)\times\sigma^2_y\\
    &= \gamma_1^2 \times (1-\eta_d^2) + \gamma_2^2 + (1-\eta^2_y)\times\sigma^2_y\\
    &= \rho^2{\frac{\eta^2_y \sigma^2_y}{1-\eta^2_d}}(1-\eta_d^2) +  (1-\rho^2) \eta^2_{y} \sigma^2_y + (1-\eta^2_y)\times\sigma^2_y\\
    &= \rho^2\eta^2_y \sigma^2_y +  (1-\rho^2) \eta^2_{y} \sigma^2_y + (1-\eta^2_y)\times\sigma^2_y\\
    &= \sigma^2_y.
\end{align*}
Therefore, the marginal distribution of $Y\mid D, X$ matches that of the observed data. We now show that we attain the desired sensitivity parameters.

First note that:
$$
\eta^2_{D\sim U\mid X} = 1- \frac{\Ep[\Var(D\mid X,U)]}{\Ep[\Var(D\mid X)]} = 1 - \frac{(1-\eta^2_d) \sigma^2_d}{\sigma^2_d} = \eta^2_d.
$$
Also, 
$$
\eta^2_{Y\sim U\mid D, X} = 1- \frac{\Ep[\Var(Y\mid D,X,U)]}{\Ep[\Var(Y\mid D, X)]} = 1 - \frac{(1-\eta^2_y) \sigma^2_y}{\sigma^2_y} = \eta^2_y.
$$
Finally, since, 
$$
\theta - \theta_s = \rho\sqrt{\frac{\eta^2_y\eta^2_d}{1-\eta^2_d}  \frac{\sigma^2_y}{\sigma^2_d}}, 
$$
by Theorem~\ref{thm:ovb-plm} we must have $Cor(g-g_s, \alpha - \alpha_s) = \rho$. \qed

\subsection{Proof of Lemma \ref{Lemma:RR}.}  
We have from the Riesz-Frechet theory presented in Appendix~\ref{app:prelim} and in particular by a direct application of the Riesz-Frechet representation theorem that given the assumptions of the Lemma, there exist unique functions $\alpha\in \Gamma$, $\alpha_s\in \Gamma_s$ such that:
\begin{align*}
\forall \gamma \in \Gamma: \Ep m(W, \gamma) =~&  \Ep \gamma(W) \alpha(W), &
\forall \gamma_s \in \Gamma_s: \Ep m(W, \gamma_s) =~& \Ep m(W_s, \gamma_s) = \Ep \gamma_s(W) \alpha_s(W), 
\end{align*}

Next we show that $\alpha_s$ is given by a projection of $\alpha$ onto $\Gamma_s$. Since $\Gamma_s\subseteq \Gamma$, we have that $\alpha$ represents the functional over $\Gamma_s$ but it is not itself in $\Gamma_s$. Consider the decomposition of $\alpha$ in the orthogonal projection $\beta_s$ of $\alpha$ onto $\Gamma_s$ and the residual $e$:
$$
\alpha(W) = \beta_s(W_s) + e(W), \quad \Ep e(W) \gamma_s(W_s) =0,  \text{ for all $\gamma_s$  in } \Gamma_s.
$$
Then we have that for all $\gamma_s\in \Gamma_s$
$$
 \Ep m(W, \gamma_s) = \Ep \gamma_s(W_s) \alpha(W) =  \Ep \gamma_s(W_s) \beta_s(W_s) + \Ep \gamma_s(W_s) e(W) =  \Ep \gamma_s(W_s) \beta_s(W_s).
$$
That is, the orthogonal projection $\beta_s$ of $\alpha$ onto $\Gamma_s$ is a RR of the functional over functions in $\Gamma_s$. Moreover, by the RF theory the RR is unique in $\Gamma_s$. By uniqueness of the RR over $\Gamma_s$, we must have that the projected $\beta_s$ coincides with $\alpha_s$ and therefore $\alpha_s$ is the orthogonal projection of $\alpha$ onto $\Gamma_s$. The latter means that we can decompose:
$$
\Ep ( \alpha - \alpha_s)^2 = \Ep \alpha^2 - 2\Ep \alpha \alpha_s + \Ep \alpha^2_s=\Ep \alpha^2 - \Ep \alpha_s^2 - 2\Ep (\alpha - \alpha_s) \alpha_s=\Ep \alpha^2 - \Ep \alpha_s^2. 
$$
where we used the fact that $\alpha_s$ is the orthogonal projection of $\alpha$ onto $\Gamma_s$ and that $\alpha_s\in \Gamma_s$, which implies that $\Ep (\alpha-\alpha_s)\alpha_s=0$.\qed
\

\subsection{Relationship of Restricted RRs of the ATE in PLM}
\label{app:plm}

Note that, as per Lemma~\ref{Lemma:RR}, $\alpha_s$ is the partially linear projection of $\alpha$ on $W_s$. Consider the  minimization problem in the partially linear regression:
$$
\min_{\beta, h} \Ep[(\alpha - \beta D - h(X))^2] 
$$
First, note that, for any $\beta$, the optimal function $h$ must equal
$$
h^*(X;\beta) = \Ep[\alpha - \beta D\mid X] = -\beta \Ep[D\mid X].
$$ 
Plugging that back into the minimization problem we now have,
$$
\min_{\beta} E\left[ \left(\alpha - \beta (D-\Ep[D\mid X]) \right)^2 \right]. 
$$
The solution for $\beta$ is then,
$$
\beta^* = \frac{\Ep[\alpha (D-\Ep[D\mid X])]}{\Ep[(D-\Ep[D\mid X])^2]} = \frac{1}{\Ep[(D-\Ep[D\mid X])^2]}.
$$
Thus,
$$
\beta^* D + h^*(X;\beta^*) = \frac{D}{E[(D-E[D\mid X])^2]} - \frac{E[D\mid X]}{E[(D-E[D\mid X])^2]} = \frac{D-E[D\mid X]}{E[(D-E[D\mid X])^2]} = \alpha_s. \qed
$$
\
\

\subsection{Proof of Lemma \ref{lemma:RR restricted}.}  
We have from the Riesz-Frechet theory presented in Appendix~\ref{app:prelim} and in particular by a direct application of the Riesz-Frechet representation theorem that given the assumptions of the Lemma, there exists a unique $\bar\alpha\in L^2(P_W)$, such that:
$$
\forall f \in L^2(P_W): \Ep m(W, f) =  \Ep f(W) \bar\alpha(W),
$$
Since, $\Gamma \subseteq L^2(P_W)$, the latter implies that:
$$
\forall \gamma\in \Gamma: \Ep m(W, \gamma) =  \Ep \gamma(W) \bar\alpha(W),
$$
that is the RR $\bar \alpha$ continues to represent the functional over the restricted linear subspace $\Gamma \subset L^2(P_W)$.  Decompose $\bar \alpha$ in the orthogonal projection $\alpha\in \Gamma$ and the residual $e$:
$$
\bar\alpha(W) = \alpha(W) + e(W), \quad \Ep e(W) \gamma (W) =0,  \text{ for all $\gamma$  in } \Gamma.
$$
Then we have that for all $\gamma\in \Gamma$
$$
\Ep m(W;,\gamma) = \Ep \gamma(W) \bar\alpha(W) =  \Ep \gamma(W) \alpha (W) + \Ep \gamma(W) e(W) =  \Ep \gamma(W) \alpha (W). 
$$
That is, $\alpha$ is a RR of the functional over functions in the space $\Gamma$, and it is unique in $\Gamma$ by the RF theory. We also have that $\Ep \bar \alpha^2 = \Ep \alpha^2 + \Ep e^2$, establishing that $\Ep \bar \alpha^2 \geq \Ep \alpha^2$.

Analogous argument yields the result for the closed linear subsets  $\Gamma_s$ of $L^2(P_{W_s})$.  \qed
\

{\subsection{Proof of Theorem \ref{thm:ovb-general}}
We decompose $\Gamma$ into $\Gamma_s\subseteq \Gamma \cap L^2(P_{W_s})$ and its orthocomplement
$\Gamma_s^{\perp}$,
$$
\Gamma=\Gamma_s\oplus\Gamma_s^\perp,
$$
so that any element $m_s \in \Gamma_s$ is orthogonal to any  $e \in \Gamma_s^{\perp}$ in the sense that
$$
\Ep m_s(W_s) e(W) = 0.
$$
By Lemma~\ref{Lemma:RR}, we can write:
$$
    \theta =~ \Ep g(W) \alpha(W), \quad
    \theta_s =~ \Ep g_s(W_s) \alpha_s(W_s)
$$
where $\alpha\in \Gamma$ and $\alpha_s\in \Gamma_s$ satisfy the properties stated in the Lemma~\ref{Lemma:RR}. In particular, $\alpha_s$ is the orthogonal projection of $\alpha$ onto $\Gamma_s$. Therefore, by definition we have that $\alpha-\alpha_s \in \Gamma_s^\perp$, i.e.,
$$\forall \gamma_s\in \Gamma_s: \Ep(\alpha(W)-\alpha_s(W_s))\gamma_s(W_s)=0.$$
Moreover, we can also argue that $g-g_s\in \Gamma_s^\perp$. By the definition of $g$ and $g_s$, and the fact that both $\Gamma$ and $\Gamma_s$ are closed linear subspaces, we have that:
\begin{align}
    \forall \gamma \in \Gamma: \Ep (Y -g(W)) \gamma(W)=0\\
    \forall \gamma_s \in \Gamma_s: \Ep (Y - g_s(W_s)) \gamma_s(W_s) = 0
\end{align}
Since $\Gamma_s\subseteq \Gamma$, the first condition also implies that:
\begin{align}
    \forall \gamma_s \in \Gamma_s: \Ep (Y -g(W)) \gamma_s(W_s)=0
\end{align}
Subtracting the last two conditions, we get:
\begin{align}
    \forall \gamma_s \in \Gamma_s: \Ep (g(W) -g_s(W_s)) \gamma_s(W_s)=0
\end{align}
which by definition implies that $g-g_s \in \Gamma_s^\perp$.
The claim of the theorem follows from 
\begin{eqnarray*}
\theta - \theta_s &=& \Ep g \alpha - \Ep g_s \alpha_s\\
&=& \Ep (g_s + g- g_s) (\alpha_s + \alpha - 
\alpha_s) - \Ep g_s \alpha_s \\
&=& \Ep g_s (\alpha - \alpha_s) +
\Ep \alpha_s (g - g_s) + \Ep (g-g_s) (\alpha-\alpha_s) \\
&=& \Ep (g-g_s) (\alpha-\alpha_s),
\end{eqnarray*}
using the fact that $\alpha_s \in \Gamma_s$ is orthogonal to $g- g_s \in \Gamma_s^{\perp}$
and $g_s \in \Gamma_s$ is orthogonal to $\alpha-\alpha_s \in \Gamma_s^{\perp}$.
By the same argument as in the proof of Theorem~\ref{thm:ovb-plm}, we also obtain,
$$
\Ep(g - g_s)(\alpha - \alpha_s) = \frac{\Ep(g - g_s)(\alpha - \alpha_s) }{\sqrt{\Ep(g - g_s)^2\Ep(\alpha - \alpha_s)^2} } \times \sqrt{\Ep(g - g_s)^2 \Ep(\alpha - \alpha_s)^2}  = \rho B,
$$
where
$$
\rho:= \frac{\Ep(g - g_s)(\alpha - \alpha_s) }{\sqrt{\Ep(g - g_s)^2\Ep(\alpha - \alpha_s)^2} } , \quad
B := {\sqrt{\Ep(g - g_s)^2\Ep(\alpha - \alpha_s)^2} }.
$$

Note $\rho \in [-1,1]$. When $\Ep(g-g_s) = \Ep(\alpha-\alpha_s)=0$, we also have $\rho = \mathrm{Cor}(g-g_s, \alpha-\alpha_s)$.

Then, using the fact that  
$$
\frac{\Ep( g- g_s)^2}{\Ep (Y- g_s)^2} = R^2_{Y-g_s \sim g- g_s},
$$
and
$$
\frac{\Ep( \alpha- \alpha_s)^2}{\Ep \alpha_s^2} =
\frac{\Ep \alpha^2-  \Ep \alpha^2_s}{\Ep \alpha_s^2}
= \frac{\Ep \alpha^2-  \Ep \alpha^2_s}{\Ep \alpha^2} \frac{\Ep \alpha^2}{\Ep \alpha_s^2} = \frac{1- R^2_{\alpha \sim \alpha_s}}{ R^2_{\alpha \sim \alpha_s}},
$$
we obtain the final expression of the OVB formula.
$\qed$
}

\subsection{Proof of Theorem ~\ref{thm:msm-ovb}}

In this section we use the following shorthand notations: $\pi = P(D=1\mid X,U)$, $\pi_s:=P(D=1\mid X)$, $O := \pi/(1-\pi)$, $O_s:= \pi_s/(1-\pi_s)$, $V:= \pi \times (1-\pi)$, $V_s:= \pi_s\times (1-\pi_s)$, $p:=P(D=1)=E[\pi_s]$. When conditioning on the event $X=x$, we use $\pi_s(x):= P(D=1\mid X=x)$, $O_s(x):= \pi_s(x)/(1-\pi_s(x))$ and $V_s(x):=\pi_s(x)\times(1-\pi_s(x))$.

Before proving the theorem, the following lemmas will be useful.

\begin{Lemma}[Bounds on the long propensity score]
\label{lemma:msm-bounded}
Under the marginal sensitivity model, conditionally on $X=x$, $\pi$ is a bounded random variable, with support on $[\pi_{\ell}(x), \pi_{u}(x)]$, where 
$$
\pi_{\ell}(x) = \frac{\Lambda^{-1}O_s(x)}{1+\Lambda^{-1}O_s(x)},\qquad \pi_u(x) = \frac{\Lambda O_s(x)}{1+\Lambda O_s(x)}.
$$
Moreover, $\Ep[\pi\mid X= x] = \pi_s(x)$.
\end{Lemma}
\begin{proof}
Recall the marginal sensitivity model imposes $\Lambda^{-1} O_s(x) \leq O(x,u) \leq \Lambda O_s(x)$. Use the fact that $O(x,u) = \pi(x,u)/(1-\pi(x, u))$ to obtain the lower and upper bounds $\pi_\ell$ and $\pi_u$. The law of iterated expectations gives $\Ep[\pi\mid X= x] = \pi_s(x)$.
\end{proof}

The next lemma appears in \cite{madansky1959bounds}, who credits \cite{Edmundson1957Bounds} for first deriving it.

\begin{Lemma}[Edmundson-Madansky Bounds] 
\label{lemma:max-of-convex}
Let $Z$ be a bounded random variable with support on $[a,b]$ and mean $\Ep Z=\mu$. Then, for convex function $f$,
$$
\Ep[f(Z)] \leq \omega f(a) + (1-\omega)f(b), \quad \text{with} \quad \omega = \frac{b-\mu}{b-a}.
$$
Moreover, the bound is sharp. If $f$ is concave, the inequality is reversed.
\end{Lemma}
\begin{proof} For any $z\in [a,b]$ we can write $z=\omega(z)a + (1-\omega(z))b$, where $\omega(z) = \frac{b-z}{b-a}$. Convexity of $f$ gives $f(z) \leq \omega(z)f(a) + (1-\omega(z))f(b).$ Applying this to the random variable and taking the expectation yields the inequality. To verify that the bound is sharp, set $Z$ to take values on $\{a, b\}$, with $P(Z=a) = \omega$, then note that $\Ep[f(Z)] =  \omega f(a) + (1-\omega)f(b)$ and $\Ep Z=\mu$. For the concave case, apply the result to $-f$.
\end{proof}

We can now combine Lemmas~\ref{lemma:msm-bounded}~and~\ref{lemma:max-of-convex} to obtain sharp bounds on any convex (or concave) function of the long propensity score, under the MSM.

\begin{Lemma}[Sharp bounds for convex $f$ under MSM]
\label{lemma:bounds-msm}
Under the marginal sensitivity model, for any convex function $f$,
$$
\Ep[f(\pi)\mid X=x] \leq \omega(x) f(\pi_{\ell}(x)) + (1-\omega)f(\pi_u(x)),
$$
where,
$$
 \omega(x) = \frac{\Lambda + O_s(x)}{(1+\Lambda)(1+O_s(x))} = \frac{\pi_s(x) + \Lambda (1-\pi_s(x)) }{ 1 + \Lambda}.
$$
Moreover, the bound is sharp. If $f$ is concave, the inequality is reversed.
\end{Lemma}

\begin{proof}
By Lemma~\ref{lemma:msm-bounded}, $\pi\mid X=x$ is a bounded random variable with support on $[\pi_{\ell}(x), \pi_u(x)]$ and mean $\pi_s(x)$. Lemma~\ref{lemma:max-of-convex} then yields the result. 
\end{proof}

We are now ready to prove the theorem. To recall, we want to show that, under the MSM,
\begin{align}
   1-R^2_{\alpha_{ATT} \sim \alpha_{ATT,s}} 
   &= 1 - \frac{\Ep O_s }{\Ep O} 
   \leq \frac{{(\Lambda -1)^2}\Ep O_s - {(\Lambda -1)^2}p}{({(\Lambda -1)^2} + \Lambda)\Ep O_s - {(\Lambda -1)^2}p} 
   \leq \frac{(\Lambda -1)^2}{(\Lambda -1)^2 + \Lambda},\nonumber\\
1-R^2_{\alpha_{ATE}\sim \alpha_{ATE,s}} 
&= 1 - \frac{\Ep[1/V_s]}{\Ep[1/V]} 
\leq \frac{{(\Lambda -1)^2}\Ep[1/V_s] - 3{(\Lambda -1)^2}}{({(\Lambda -1)^2} + \Lambda)\Ep[1/V_s] - 3{(\Lambda -1)^2}} 
\leq \frac{(\Lambda -1)^2}{(\Lambda -1)^2 + \Lambda},\nonumber\\
1-R^2_{\alpha_{PLM} \sim \alpha_{PLM,s}} 
&= 1 - \frac{\Ep V}{\Ep V_s} 
\leq  1-\frac{\Ep\left[\frac{\Lambda V_s}{\Lambda + (\Lambda -1)^2V_s}\right]}{\Ep V_s}
\leq \frac{(\Lambda -1)^2}{(\Lambda -1)^2 + 4\Lambda}.\nonumber
\end{align}
The first bound is sharp given $\Lambda$ and observed data. The second bound is sharp given $\Lambda$ alone.

\begin{proof}[Proof of Theorem~\ref{thm:msm-ovb}]

Our proof strategy is the following. We first use Lemma~\ref{lemma:bounds-msm} to obtain sharp bounds that hold conditionally for any $X=x$. Then the marginal bounds  follow from taking expectations. The data-independent version of the bound is obtained by taking the supremum over any possible realization of the observed data. 

\textbf{Gain in Odds.} We start with the gain in average odds. Note the odds function is convex. Applying Lemma~\ref{lemma:bounds-msm} yields the sharp bound,
\begin{align*}
\Ep[O\mid X=x] &\leq \omega(x) \pi_\ell(x)/(1-\pi_\ell(x)) + (1-\omega(x)) \pi_u(x)/(1-\pi_u(x)) \\
&= \omega(x) \Lambda^{-1}O_s(x) + (1-\omega(x))\Lambda O_s(x) \\
&= (\omega(x) \Lambda^{-1} + (1-\omega(x))\Lambda) O_s(x).
\end{align*}

Using the identities $\omega(x) \Lambda^{-1} + (1-\omega(x))\Lambda = 1 + \left(\frac{(\Lambda - 1)^2}{\Lambda}\right)\pi_s(x)$ and $O_s(x)\pi_s(x) = O_s(x) -\pi_s(x)$ gives us
$$
\Ep[O\mid X=x]\leq \left(\frac{\Lambda +(\Lambda -1)^2}{\Lambda}\right) O_s(x) -\left(\frac{(\Lambda -1)^2}{\Lambda}\right)\pi_s(x).
$$
Applying this to the random variable, taking expectations and rearranging yields the data-dependent bound. To obtain the data-independent bound, take $\Ep O_s \to \infty$.

\textbf{Gain in Precision.} We now move to the gains in average precision. First note we can write the inverse variance as
$$
\frac{1}{V} = O + \frac{1}{O} + 2.
$$
By symmetry, the previous bound holds for $\Ep[1/O\mid X=x]$, replacing $O_s(x)$ with $1/O_s(x)$ and $\pi_s$ with $1-\pi_s$. Then,
$$
\Ep[1/V\mid X=x]\leq   \left(\frac{\Lambda +(\Lambda -1)^2}{\Lambda}\right) O_s(x) + \left(\frac{\Lambda +(\Lambda -1)^2}{\Lambda}\right) 1/O_s(x) - \frac{(\Lambda -1)^2}{\Lambda} + 2
$$
Rearranging terms, and using again the identity $1/V_s(x) = O_s(x) + 1/O_s(x) +2$  we can simplify the last expression to,
$$
\Ep[1/V\mid X=x]\leq   \left(\frac{\Lambda + (\Lambda -1)^2}{\Lambda} \right)(1/V_s(x)) - 3\left(\frac{(\Lambda -1)^2}{\Lambda}\right).
$$
Applying this to the random variable,  taking expectations and rearranging yields the data-dependent bound. To obtain the data-independent bound, take $\Ep[1/V_s] \to \infty$.

\textbf{Gain in Variance Explained.} Moving to the non-parametric partial $R^2$, note that the variance function is concave. We thus have the sharp \emph{lower} bound,
\begin{align*}
\Ep[V\mid X=x] &\geq \omega(x) \times \pi_\ell(x)\times (1-\pi_\ell(x)) + (1-\omega(x))\times \pi_u(x) \times (1-\pi_u(x))\\
&= \frac{\Lambda O_s(x)}{(\Lambda + O_s(x))(1+\Lambda O_s(x))}.
\end{align*}
Divide the numerator and denominator by $(1+O_s(x))^2$ and use the identities $V_s(x) = O_s(x)/(1+O_s(x))^2$, $(\Lambda + O_s(x))(1+\Lambda O_s(x)) = \Lambda(1+O_s(x))^2 + O_s(x)(\Lambda -1)^2$ to obtain,
$$
\Ep[V\mid X=x] \geq \frac{\Lambda V_s(x)}{\Lambda + (\Lambda - 1)^2V_s(x)}.
$$
Applying this to the random variable,  taking expectations and rearranging yields the data-dependent bound. To obtain the data-independent bound, take $V_s(x) \to 1/4$. 

\textbf{Bounds on $C_D$.} Bounds on $C_D$ can be obtained via the mapping $C^2_D = (1-R^2_{\alpha \sim \alpha_s})/R^2_{\alpha \sim \alpha_s}$.
\end{proof}

\begin{remark} These bounds are in fact reached by the adversarial confounder of the construction in \cite{dorn2023sharp}, when proving sharpness of their MSM bounds for the average potential outcome, ATE and ATT.
\end{remark}

\subsection{Rosenbaum's model and MSM can have unbounded parameters}
\label{app:odds}

Recall that the marginal sensitivity model posits $\Lambda \geq 1$ such that
$$
\frac{1}{\Lambda} \leq \frac{\Odds(D\mid X=x,U=u)}{\Odds(D\mid X=x)} \leq \Lambda, \qquad \forall x \in \mathcal{X}, u \in \mathcal{U}.
$$
Similarly, Rosenbaum's model posits a $\Gamma \geq 1$ such that
$$
\frac{1}{\Gamma} \leq  \frac{\Odds(D\mid X=x,U=u)}{\Odds(D\mid X=x, U = u')} \leq \Gamma, \qquad \forall x \in \mathcal{X},  u \in \mathcal{U}.
$$
We now show that these two parameters can be unbounded while the OVB is actually small. Let the unobserved confounder $U$ be normally distributed and consider the case with no observed covariates~$X$. Let the full propensity score be
\begin{align}
    P(D=1\mid U = u) &= \frac{e^{\rho u}}{1+ e^{\rho u}}.
\end{align}
We then have that $\Odds(D\mid X=x,U=u)/\Odds(D\mid X=x,U=u')= e^{\rho (u-u')}$ and $\Odds(D\mid X=x,U=u)/\Odds(D\mid X=x)= e^{\rho u}$. 

Thus, the true sensitivity parameters $\Gamma$ and $\Lambda$ are unbounded, $$\Gamma = \Lambda = \infty,$$ once $\rho \neq 0$. In contrast, the true OVB parameter $1-R^2_{\alpha \sim \alpha_s}$ (either for the ATE or ATT) converges to 0 as $\rho \to 0$.  For example, with $\rho =.1$, the worst-case odds ratio is infinite, whereas the true OVB bound is very tight, since we have  $1-R^2_{\alpha_{ATE}\sim \alpha_{ATE,s}} \approx  0.0025$ and $1-R^2_{\alpha_{ATT} \sim \alpha_{ATT,s}} \approx 0.005$. 

In summary, in this example, the true sensitivity parameters translate into tight bounds on the ATE or ATT in our OVB-based approach, versus uninformative bounds in worst-case odds-ratio approaches.  We note that similar discussion applies to approaches that constrain worst-case risk-ratios, such as those in \cite{ding:ecm2016}.

\subsection{Proof of Lemma \ref{lemma:DML} and Theorem
\ref{thm:DML}}
The Lemma follows from the application of Theorem 3.1 and Theorem 3.2 in \cite{DML}.  Valid estimation of covariance follows similarly to the proof of Theorem 3.2 in \cite{DML}. The first result of Theorem \ref{thm:DML} follows from the delta method in \cite{vdV-W}. The validity of the confidence intervals follows from using the standard arguments for confidence intervals based on asymptotic normality.\qed

\subsection{Proof of Theorem \ref{thm:ovb-validity}} Here the argument is similar to 
 \cite{RR-local}, but we provide details for completeness. 

The assumptions directly imply that the candidate long RR  obey $\alpha \in L^2(P)$ with $\|\alpha\|_{P,2} \leq C$ in each of the examples, for some constant $C$ that depends on $P$. By $\Ep Y^2 < \infty$, we have $g \in L^2(P)$. Therefore, $| \Ep \alpha g| < \| \alpha \|_{P,2} \| g\|_{P,2}< \infty$ in any of the calculations below.

We first verify that long RR $\alpha$'s can indeed represent the functionals $g \mapsto \theta(g):= \Ep m(W, g)$ in Examples 1,2,3,5 over $g \in L^2(P)$. In Example 4, the long RR represents the Hahn-Banach extension of the mapping $g \mapsto \theta(g)$ to $L^2(P)$ over $L^2(P)$.

In Example 1, recall that
$\bar \ell (X,U) := \Ep [\ell(W_s)\mid X,U]$. Then since $dP(d,x,u) =  \sum_{j=0}^1 1(j=d) P[D=j\mid X=x, U=u]  dP(x,u)$ by  Bayes rule, we have
\begin{align*}
\Ep g(W) \alpha(W) 
& =  \int g (d, x, u) \frac{1(d=\bar d)\bar \ell(x,u)}{P[D= \bar d \mid  X=x, U=u]} dP(d,x,u) =  \int g(\bar d, x, u) \bar \ell(x,u) d P(x,u) \\
&=  \Ep g(\bar d, X, U) \bar \ell(X,U) = \Ep g(\bar d, X, U)  \ell(W_s)= \theta(g),    
\end{align*}
where the penultimate equality follows by the law of iterated expectations. The claim for Example 2 follows from the claim for Example 1.
 
Example 3 follows by the change of measure of $dP_{\tilde W}$ to $dP_{W}$, given the assumed  absolutely continuity of the former with respect to the latter. Then we have
\begin{align*}
\Ep g(W) \alpha(W) 
&=  \int g
\left(\Ep[\ell \mid \tilde W]\frac{dP_{\tilde W}}{dP_W} - \ell\right)  dP_W \\
&=  \int  g \Ep[ \ell \mid \tilde W]\, dP_{\tilde W} - \int (g \ell) dP_W \\
&=  \Ep \left[g(\tilde W)\Ep[\ell(W_s)\mid \tilde W]\right] - \Ep[g(W)\ell(W_s)] \\
&=  \Ep[g(\tilde W) \ell(W_s)] - \Ep[g(W)\ell(W_s)] \\
&=  \Ep[\{g( T(W_s), U) - g(W_s, U)\}\ell(W_s)]  = \theta(g).
\end{align*}
This also corrects the expression for Example 3 in \cite{RR-local}.

In Example 4, we can write for any $g$ that has the properties stated in this example:
\begin{eqnarray*}
\Ep g(W) \alpha(W) & =&  - \int \int g(w) \frac{\divg_d ( \ell(W_s) t(W_s) f(d\mid x,u))}{f(d\mid x,u)} f(d\mid x,u)\mathrm{d} d \mathrm{d}P(x,u) \\
&=&  - \int \int g(w) \divg_d ( \ell(W_s) t(W_s) f(d\mid x,u)) \mathrm{d} d \mathrm{d}P(x,u)\\
& =&  -\int \int_{\partial \mathcal{D}_{u,x}}  
g(w) t(W_s)' \ell(W_s) f(d\mid x,u) n_{u,x}(u) \mathrm{d} S(d) \mathrm{d} P(x,u) \\
& & + 
 \int \int  \partial_d g(w) ' t(W_s) \ell(W_s) f(d\mid x,u) \mathrm{d} d \mathrm{d} P(x,u) \\
& =&  \ \int \int  \partial_d g(w) ' t(W_s) \ell(W_s) f(d\mid x,u) \mathrm{d} d \mathrm{d} P(x,u) = \theta(g),
\end{eqnarray*}
where we used the integration by parts and that  $\ell(W_s) t(W_s) f(d\mid x,u)$ vanishes for any $d$  in the boundary of $\mathcal{D}_{x,u}$.

Example 5 follows by the change of measure $d P_U \times d F_k$ to $dP_{W}$, given the assumed  absolutely continuity of the former with respect to the latter on $\mathcal{U} \times \text{support} (\ell)$.  Then we have
\begin{eqnarray*}
\Ep g(W) \alpha(W) &=& \int g \ell  
\left (\frac{ [ dP_U \times d (F_1 - F_0)]}{d P_{W}} \right)  dP_W \\
&= & \int g(W_s, u)\ell(W_s) dP_U(u) d (F_1 - F_0)(W_s) = \theta(g).
\end{eqnarray*}

In all examples, the continuity of $g \mapsto \theta(g)$ required in Assumption 1 now follows from the representation property and from $|\Ep \alpha g| \leq \| \alpha\|_{P,2} \| g\|_{P,2} \leq C \| g\|_{P,2}$.

Verification of Assumption 2 follows directly from the inspection of the scores given in Section 5. 

Note that we do not need the analytical form of the short RRs to verify Assumptions 1 or 2. However, their analytical form can be found by exactly the same steps as above, or by taking the conditional expectation. \qed

\subsection{Discussion on sharpness}
\label{app:sharpness}

Here we note that certain model assumptions in combination with the distribution of observed data $P$ can place restrictions on the admissible values of sensitivity parameters. Recall the omitted variable bias formula is,
$$
\theta - \theta_s = \rho \times \sqrt{R^2_{Y- g_s \sim g- g_s} \times \left(\frac{1-R^2_{\alpha\sim \alpha_s}}{R^2_{\alpha\sim \alpha_s}}\right)}\times S.
$$
Given plausibility judgments on the maximum values of the components of the bias formula, i.e., $|\rho| \leq \rho^{\max}$, $R^2_{Y- g_s \sim g- g_s}\leq R^{2\max}_Y$ and $1-R^2_{\alpha\sim \alpha_s} \leq R^{2\max}_D$, the OVB formula immediately implies the  bound,
$$
|\theta - \theta_s| \leq \rho^{\max} \times \sqrt{R^{2\max}_{Y} \times \left(\frac{R^{2\max}_{D}}{1-R^{2\max}_{D}}\right)}\times S.
$$
Formally we say this bound is sharp given model assumptions $\mathcal{M}$, and observed data $\mathcal{P}$, if, for any triplet $(\rho^{\max},R^{2\max}_{Y}, R^{2\max}_{D}) \in [0,1)^3$, we can construct a full data law such that: (a) it is compatible with $\mathcal{M}$ and $\mathcal{P}$; and, (b) achieves (or gets arbitrarily close to) $|\rho| = \rho^{\max}$, $R^2_{Y- g_s \sim g- g_s}= R^{2\max}_Y$ and $1-R^2_{\alpha\sim \alpha_s} =R^{2\max}_D$. In general, one takes  $\mathcal{P}$ to be the full joint distribution of the observed data.

It is easy to see that the OVB bound may not be sharp in certain cases, depending on $\mathcal{M}$ and $\mathcal{P}$. One example is the case of the ATE when the outcome is bounded. If, say $ 0 \leq Y \leq 1$, the bias $|\theta - \theta_s|$ itself is also naturally bounded. This means that the sensitivity parameters are not variation independent. When that is the case, {some} combinations of the sensitivity parameters (especially when $R^{2\max}_{D}$ is high enough), may not be compatible with the observed data, meaning that the bound is conservative. However, note the bound could still be sharp for {most} plausible combinations of the sensitivity parameters, even in such scenarios. Having said that, a simple solution to tighten the bounds in such scenarios is to cap the OVB formula at the ``no assumptions'' bounds, whenever these are known (e.g. Manski's bounds). This approach always yields valid bounds that are at least as informative as the unconstrained OVB formula.

More generally, our OVB framework generates bias formulas for a wide range of target parameters and model specifications. Given this breadth, it is difficult to establish general sharpness claims that apply broadly across many functionals, models, and data constraints. We believe specialized results and case-by-case analyses will be needed as demand for such results grows. The example in Appendix~\ref{app:sharp-plm} illustrates how one can derive such results, and we leave a broader characterization of sharpness to future work. Finally, we note that these characterizations do not preclude the usefulness of the bounds. Many widely used sensitivity analysis methods lacked general sharp results for years, or even {decades}, after their initial proposals. Indeed  the sharpness results of \cite{dorn2023sharp} and \cite{dorn2025doubly} were established almost two decades after \cite{tan:jasa2006}.

\section{Extended Literature Review}
\label{sec:lit}
We now provide a more extended discussion of the related literature on sensitivity analysis.  We focus the discussion on recent methods, and on how they differ from our proposal. We refer readers to \cite{liu:ps2013}, \cite{richardson:ss2014}, \cite{CH2020}, and \cite{scharfstein2021semiparametric} for further details.

In contrast to our approach, many of the earlier works on sensitivity analyses demand from users a rather extensive specification, or parameterization, of the nature of unobserved confounders. This could range from positing the marginal (or conditional) distribution of these latent variables, along with specifying how such confounders would enter the outcome or treatment equations (e.g., entering linearly). Among such proposals, with varying degrees of requirements and parametric assumptions, we can find, e.g., \cite{rosenbaum1983assessing}, \cite{imbens2003sensitivity}, \cite{dorie2016flexible}, \cite{altonji2005selection}, \cite{cinelli:icml2019},\cite{zhang2021exploiting}. 

Another branch of the sensitivity literature requires users to specify 
instead  a ``tilting,'' ``selection,'' or ``bias'' function, directly parameterizing the difference between the conditional distribution of the outcome under treatment (control) between treated and control units; or, when the target parameter is the ATE, just parameterizing the difference in conditional means. Earlier work on this area goes back to \cite{robins1999association} and \cite{brumback2004sensitivity} with more recent works from  \cite{franks2020flexible} and \cite{scharfstein2021semiparametric}, the latter with a special focus on binary treatments, and flexible semi-parametric estimation procedures. Our proposal differs from this literature in that we do not model the bias directly, instead we impose constraints on the maximum explanatory power of confounders.

Continuing with binary treatments, many sensitivity proposals focus on this special case. They differ mainly on how to parameterize departures from random assignment. For instance, \cite{masten:e2018} places bounds on the \emph{difference} between the treatment assignment distribution, conditioning and not conditioning on potential outcomes, whereas \cite{rosenbaum:b1987} and more recently \cite{tan:jasa2006, zhao:jrssb2019, yadlowsky2022bounds} place bounds on the worst-case \emph{odds} of such distributions.  \cite{bonvini2021sensitivity}, on the other hand, propose a contamination model approach, placing restriction on the \emph{proportion of confounded units}.  Our approach is different from all these approaches in at least two main ways. First, we do not restrict our analyses to the binary treatment case. Second, even in the important case of a binary treatment, we parameterize violations of ignorability via the exact OVB formula for each estimand (see discussion in Section~\ref{sec:msm}), resulting in different sensitivity parameters, such as the \emph{gains in average precision} for the ATE, or \emph{gains in average odds} for the ATT. Our sensitivity parameters are thus different from these approaches (we provide a numerical example in Appendix~\ref{app:odds}, which demonstrates practical and theoretical value of the new parameterization).

Other sensitivity results, while allowing for general confounders, treatments and outcomes, restrict their attention to  specific target parameters. For example, 
\cite{ding:ecm2016} derive general bounds for the risk-ratio, with sensitivity parameters also in terms of risk-ratios. Our approach is thus different both in terms of target parameters (continuous linear functionals of the CEF), and in terms of sensitivity parameters ($R^2$ based sensitivity parameters). \cite{CH2020} derive bounds for linear regression coefficients.
Their result is a special case of ours when the target functional is the coefficient of a linear projection.  Their approach does \textit{not} cover nonlinear regression and the causal parameters that we study here (e.g., it does not cover the ATE in the nonparametric model with a binary~treatment). Finally, \cite{deto2021} provide an alternative expression for omitted variable bias of average causal derivatives.

\section{Benchmarking Analysis}
\label{app:benchmarks}

Here we describe our new approach to benchmarking in nonparametric models. Our analysis is partly inspired by benchmarking analyses previously proposed in \cite{imbens2003sensitivity}, \cite{altonji2005selection}, \cite{oster2019unobservable}, and more recently \cite{CH2020}. In particular our proposal is closest in nature to the latter reference for linear regression, by postulating that the gains in explanatory power due to latent variables is similar to the gains in explanatory power of observed variables.

\subsection{Notation.} 
We start by setting notation. For a given observed covariate $X_j$, let $X_{-j}$ denote the set of all other observed covariates $X$ except $X_j$. Let $g_{s,-j}$ and $\alpha_{s,-j}$ denote the short regression function and the short RR \emph{excluding} covariate $X_j$. We define the gain in the explanatory power of $X_j$ with the RR as: $$
1-R^2_{\alpha_s \sim \alpha_{s,-j}} = \frac{\Ep \alpha_s^2 - \Ep \alpha_{s,-j}^2}{\Ep \alpha_s^2}.
$$
We also define the change in estimate, $\Delta \theta_{s,j} :=\Ep m(W, g_s)- \Ep m(W, g_{s,-j})$, and the correlation,
$
\rho_{j} := \mathrm{Cor}(g_s - g_{s,-j}, \alpha_s - \alpha_{s,-j}).
$

\subsection{Relative bounds on \texorpdfstring{$1-R^2_{\alpha \sim \alpha_s}$}{1-R2}}

Note we can write $1-R^2_{\alpha \sim \alpha_s}$ as,
$$
1-R_{\alpha \sim \alpha_{s}}^{2}=1-\frac{\Ep\alpha_{s}^{2}}{\Ep\alpha^{2}}.
$$
Now dividing and multiplying the fraction by $\Ep \alpha_{s,-j}^{2}$ we obtain the following decomposition:
$$
1-R_{\alpha \sim \alpha_{s}}^{2}
= 1-\frac{\Ep \alpha_{s}^{2}}{\Ep \alpha_{s,-j}^{2}} \frac{\Ep \alpha_{s,-j}^{2}}{E\alpha^{2}} =\frac{R_{\alpha_{s} \sim \alpha_{s,-j}}^{2}-R_{\alpha \sim \alpha_{s,-j}}^{2}}{R_{\alpha_{s} \sim \alpha_{s,-j}}^{2}}
=\frac{(1-R_{\alpha \sim \alpha_{s,-j}}^{2})-(1-R_{\alpha_{s} \sim \alpha_{s,-j}}^{2})}{R_{\alpha_{s} \sim \alpha_{s,-j}}^{2}}.
$$
The numerator stands for the additive gain in variation that the latent variables $U$ create  in the $\mathrm{RR}$, in addition to what $X_{j}$ creates.  We can now define the following measure $k_{D,j}$ of \emph{relative} strength of $U$, which captures how  much  $U$ adds in terms of variation explained of the RR, as compared to the observed gains due to $X_j$,
\begin{align}
\label{eq:kd}
k_{D,j} := \frac{R_{\alpha_{s} \sim \alpha_{s,-j}}^{2}- R_{\alpha \sim \alpha_{s,-j}}^{2}}{1-R_{\alpha_{s} \sim \alpha_{s,-j}}^{2}}.   
\end{align}
This allows us to rewrite the sensitivity parameter $1-R_{\alpha \sim \alpha_{s}}^{2} $ in terms of relative measure $k_{D,j}$ and the observed strength~of~$X_j$:
\begin{align}
\label{eq:bench-RR}
1-R_{\alpha \sim \alpha_{s}}^{2} = k_{D,j} \left( \frac{1-R_{\alpha_{s} \sim \alpha_{s,-j}}^{2}}{R_{\alpha_{s} \sim \alpha_{s,-j}}^{2}}\right).    
\end{align}
In a partially linear model, the above reparameterization corresponds to the following result:
\begin{align}
\label{eq:bench-RR-PLM}
1-R_{\alpha \sim \alpha_{s}}^{2} = \eta^2_{D \sim U \mid X}
= k_{D,j} \left(\frac{\eta_{D \sim X_{j} \mid X_{-j}}^{2}}{1-\eta_{D \sim X_{j} \mid X_{-j}}^{2}}\right).     
\end{align}
The usefulness of equations~(\ref{eq:bench-RR})~and~(\ref{eq:bench-RR-PLM}) is that they allow researchers to bound the bias by making claims of relative importance of $U$, as compared to $X_j$. For example, setting $k_{D,j} \leq 1$ is equivalent to claiming that the additive gains in explanatory power due to latent confounders is no greater than the observed gains in explanatory power due to $X_j$.

\subsection{Relative bounds on \texorpdfstring{$\eta^2_{Y\sim U\mid DX}$}{eta2}} Here we follow a similar strategy as in the previous section. First note we can write $\eta^2_{Y\sim U\mid DX}$ as,
\begin{align}
    \eta^2_{Y\sim U\mid DX} = \frac{\eta^2_{Y\sim UX_j\mid DX_{-j}} -  \eta^2_{Y\sim X_j\mid DX_{-j}}}{1-\eta^2_{Y\sim X_j\mid DX_{-j}}}.
\end{align}
Now define the measure of relative strength $k_{Y,j}$,
\begin{align}
\label{eq:ky}
k_{Y,j} := \frac{\eta^2_{Y\sim UX_j\mid DX_{-j}} -  \eta^2_{Y\sim X_j\mid DX_{-j}}}{\eta^2_{Y\sim X_j\mid DX_{-j}}}.
\end{align}
Note $k_{Y,j}$ stands for how much variation is explained by adding $U$ to the regression equation, as compared to the observed gains in explanatory power due to $X_j$. This allows us to rewrite $\eta^2_{Y\sim U\mid DX}$ as a function of the relative strength $k_{Y,j}$ and the observed strength of $X_j$, as in
\begin{align}
\label{eq:bench-eta}
\eta^2_{Y\sim U\mid DX} = k_{Y,j} \left(\frac{\eta^2_{Y\sim X_j\mid DX_{-j}}}{1 - \eta^2_{Y\sim X_j\mid DX_{-j}}}\right).  
\end{align}

\subsection{Benchmarking \texorpdfstring{$|\rho|$}{rho}} The correlation $\rho$ is not a measure of strength or explanatory power of the latent variables. Rather, it measures how much errors in the outcome equation are systematically related to errors in the Riesz representer. That is, for a confounder to create bias, this confounder not only needs to be strongly associated with the treatment and the outcome, but also the functional form of these associations needs to be ``similar'' in both equations, in order to create systematic biases. For instance, consider the (extreme) example discussed in the text with structural equations:
\begin{align}
D &= U^2\\
Y &= \theta D + U
\end{align}
where $U\sim N(0,1)$. Here, even though the latent variable $U$ (nonparametrically) explains 100\% of the residual variation in both the treatment and the outcome equations, the nonlinearity of the confounding model attenuates this bias, making it effectively zero, because $U^2$ is uncorrelated with~$U$.

Therefore, plausibility judgments on the magnitude of $|\rho|$ will depend on how much we expect the functional form of the latent confounder in the treatment and outcome equations to be similar. In order to calibrate this judgment from empirical data, we propose using as a reference the observed Pearson's correlation of the outcome and RR errors induced by $X_j$, as given by $\rho_j$.

\subsection{Statistical inference of benchmark components}

We have the following measure of strength of association of the confounders with the outcome:
\begin{equation}\label{eq:bench1}
\eta^2_{Y \sim U \mid DX}
= k_{Y,j} \left(\frac{\eta_{Y \sim X_{j} \mid DX_{-j}}^{2}}{1-\eta_{Y \sim X_{j} \mid DX_{-j}}^{2}} \right) =: k_{Y,j} G_{Y,j}
\end{equation}
We also have the following measure of strength of association of the confounders with the RR:
\begin{equation}\label{eq:bench2}
1-R_{\alpha \sim \alpha_{s}}^{2} = k_{D,j} \left(\frac{1-R_{\alpha_{s} \sim \alpha_{s,-j}}^{2}}{R_{\alpha_{s} \sim \alpha_{s,-j}}^{2}} \right) =: k_{D,j} G_{D,j}.
\end{equation}
The latter metric, in a partially linear model, corresponds to:
$$
1-R_{\alpha \sim \alpha_{s}}^{2} 
= \eta^2_{D \sim U \mid X} 
= k_{D,j} \left( \frac{\eta_{D \sim X_{j} \mid X_{-j}}^{2}}{1-\eta_{D \sim X_{j} \mid X_{-j}}^{2}}\right). 
$$

We call the estimable components $G_{Y,j}$ and $G_{D,j}$ above the ``gain''  metrics. They measure gains in the explanatory power of observed covariates and, under the stated hypotheses of $k_{Y,j}$ and $k_{D,j}$, serve as proxies for the sensitivity parameters $\eta^2_{Y \sim U \mid DX}$ and $1-R_{\alpha \sim \alpha_{s}}^{2}$. These quantities also immediately pin-down $C_Y^2= \eta^2_{Y \sim U \mid DX}$ and $C_D^2=(1-R_{\alpha \sim \alpha_{s}}^{2})/R_{\alpha \sim \alpha_{s}}^{2} $ that enter the bias formulas. Since these components need to be estimated from the data, we use the following debiased representations which we now discuss.

\begin{remark}[Debiased Representations] We use the following Neyman orthogonal representations for the gain metrics, 
$$
G_{Y, j} = \frac{\sigma^2_{s, -j}}{\sigma^2_{s}} - 1, \qquad G_{D, j} = \frac{\nu^2_{s}}{\nu^2_{s, -j}} - 1,
$$
where the variance of the residuals, $\sigma^2_{s}:= \Ep(Y-g_{s})^2$ and $\sigma^2_{s, -j}:= \Ep(Y-g_{s, -j})^2$, are already in debiased form, and 
$$ \nu^2_s := 2 \Ep m(W, \alpha_s) - \Ep \alpha^2_s \text{ and } \nu^2_{s,-j}:= 2 \Ep m(W, \alpha_{s,-j}) - \Ep \alpha^2_{s,-j}$$ are the debiased forms for $\Ep \alpha_s^2$ and $\Ep \alpha_{s,-j}^2$.

As for $\rho_j$, we first define the debiased form of the change in estimates,
$$\Delta \theta_{s,j} =
\Ep m(W,g_s) + \Ep  (Y-g_s) \alpha_s - 
\Ep m(W, g_{s,-j}) - \Ep  (Y - g_{s,-j}) \alpha_{s,-j}.
$$ 

This gives us the debiased representation for the correlation,
$$
\rho_{j} = \frac{\Delta \theta_{s,j}}{\sqrt{\sigma^2_{s, -j} - \sigma_s^2}\sqrt{\nu^2_s - \nu^2_{s,-j}}}.
$$
The debiasedness (Neyman orthogonality) of the above expressions follows from the chain rule for functional calculus (e.g., \cite{vdV-W}),  exploiting the fact that each representation is a smooth transformation of debiased representations. Note the components of $G_{D, j}$, $G_{Y, j}$, and $\rho_j$ are the same as those already covered by Lemma~\ref{lemma:DML}, with the minor change of dropping covariate $X_j$ where appropriate.   Given DML estimators of the components, we then obtain the following influence functions for the plug-in estimators of $G_{D, j}$, $G_{Y, j}$, and $\rho_j$:
\begin{align}
\psi^o_{G_{Y, j}}(Z) &= \frac{\sigma^2_{s} \psi^o_{\sigma^2_{s,-j}}(Z) - \sigma^2_{s,-j}  \psi^o_{\sigma_s^2}(Z) }{\sigma^4_s} \nonumber \qquad
\psi^o_{G_{D, j}}(Z) = \frac{\nu^2_{s,-j} \psi^o_{\nu_s^2}(Z) - \nu^2_{s}  \psi^o_{\nu^2_{s,-j}}(Z) }{\nu^4_{s,-j}} \nonumber \\
\psi^o_{\rho_j}(Z) &= 
\frac{\psi^o_{\theta_{s}}(Z) - \psi^o_{\theta_{s,-j}}(Z)}{(\sigma^2_{s, -j} - \sigma_s^2)^{1/2} (\nu_s^2 - \nu^2_{s,-j})^{1/2}} 
- \frac{\Delta \theta_{s, j}(\psi^o_{\sigma_{s,-j}^2}(Z) - \psi^o_{\sigma_s^2}(Z))}{2(\sigma^2_{s, -j} - \sigma_s^2)^{3/2} (\nu_s^2 - \nu^2_{s,-j})^{1/2}}  
\nonumber \\
&- \frac{\Delta \theta_{s, j}(\psi^o_{\nu_{s}^2}(Z) - \psi^o_{\nu_{s,-j}^2}(Z))}{2(\sigma^2_{s,-j} - \sigma_s^2)^{1/2} (\nu_s^2 - \nu^2_{s, -j})^{3/2}}.
\end{align}
\end{remark}

\subsection{Statistical inference for the bounds}
\label{app:inference-benchmark-bound}

Let $\text{BF}=\rho C_Y C_D$ denote the ``bias factor'' of the bounds, that is, the part of the bounds that depends on quantities related to unobserved confounder $U$. When we perform benchmarking, this bias factor is estimated based on data, and the new plug-in estimator of the bounds is given by:
$$
\widehat{\theta}_{\pm} := \widehat{\theta}_s \pm \underbrace{\left(\widehat{\rho}_j \times  \sqrt{k_{Y, j}\widehat{G}_{Y, j} \times \frac{k_{D,j} \widehat{G}_{D,j}}{1-k_{D,j} \widehat{G}_{D,j}}}\right)}_{\widehat{\text{BF}}} \times \widehat{S} = \widehat{\theta}_s \pm \widehat{\text{BF}}\times \widehat{S} 
$$
where the components $\widehat{\rho}_j$, $\widehat{G}_{D, j}$ and $\widehat{G}_{Y, j}$ are estimated using the DML algorithm and the debiased representations given above.  Estimation errors on the estimable components of BF can thus be easily propagated via the delta method.

In particular, given the product form of the bias, $\widehat{\text{BF}}\times \widehat{S}$, the adjustment in the influence function of Theorem~\ref{thm:DML} takes a simple form---it suffices to add an extra term due to the estimation error of BF, scaled by S. That is, 
$$
\varphi^o_{\pm}(Z) = \underbrace{\psi^o_{\theta_s}(Z) \pm \text{BF}\times \psi^o_S(Z)}_{\text{Influence function with fixed BF}}~\pm~ \underbrace{\psi^o_\text{BF}(Z) \times S}_{\text{Addition due to estimated BF}},
$$
where here for simplicity we denote by $\psi^o_S(Z)$ the influence function of the scaling factor $S$ as described in the main text,
$$
\psi^o_S(Z):= \frac{1}{2S} \left( \sigma^2_s \psi^o_{\nu_s^2}(Z) + \nu_{s}^2 \psi^o_{\sigma_s^2}(Z)\right),
$$
and  $\psi^o_\text{BF}(Z)$ is the influence function of the bias factor BF. This last term can be computed using the previous results:
\begin{align}
\psi^o_\text{BF}(Z) := 
& \left( \frac{k_{D,j} G_{D,j}}{1-k_{D,j} G_{D,j}} \right)^{1/2} \times  (k_{Y, j} G_{Y, j})^{1/2} \times \psi^o_{\rho_j}(Z)    \nonumber \\
%
&+ \rho_j  \times \left( \frac{k_{D,j} G_{D,j}}{1-k_{D,j} G_{D,j}} \right)^{1/2} \times  \frac{k_{Y, j}\psi^o_{G_{Y, j}}(Z)}{2 ({k_{Y, j} G_{Y, j}})^{1/2}} \nonumber\\ 
&+  \rho_j \times (k_{Y, j}G_{Y, j})^{1/2} \times 
\left( \frac{k_{D,j} \psi^o_{G_{D,j}}(Z) }{2 (k_{D, j} G_{D, j})^{1/2}(1-k_{D,j}G_{D,j})^{3/2}}\right).
\end{align}
For simplicity, we do not propagate uncertainty of these metrics into the bounds in the main text.

\subsection{Empirical Benchmarking Results for 401(k) Example.} Using the formulas described above, we obtain the following empirical results for the 401(k) example. Table~\ref{table:other x plm} shows the results for the partially linear model and Table~\ref{table:other x} shows the results for the nonparametric model.\footnote{All metrics are estimated using the same procedure described in footnote~\ref{foot:dml-fit}.}  These gain metrics are the ones used in the contour plots of the main text.
\begin{table}[t]
\begin{center}
\begin{tabular}{lrrrr}
  \hline
  \hline
  & \multicolumn{2}{c}{Gain Metrics} & Correlation & Change in estimate\\
  \cmidrule{2-5}
 Observed covariate & $G_{Y,j}$  & $G_{D,j}$ &  $\rho_{j}$ & $\Delta \widehat{\theta}_{s,j}$ \\ 
  \hline 
  inc     & 0.145 & 0.047 & 0.34  & 3,349 \\ 
  pira    & 0.038 & 0.003 & 0.21  & 188 \\ 
  twoearn & 0.021 & 0.007 & -0.25 & -621 \\ 
  \hline
   \hline
\end{tabular}
\end{center}
\small 
\vspace{-.1in}
\caption{Explanatory power of observed covariates in the Partially Linear Model. All estimates are debiased and cross-fitted. 
}
\label{table:other x plm}
\end{table}

\begin{table}[ht]
\begin{center}
\begin{tabular}{lrrrr}
  \hline
  \hline
  & \multicolumn{2}{c}{Gain Metrics} & Correlation & Change in estimate\\
  \cmidrule{2-5}
 Observed covariate & $G_{Y,j}$  & $G_{D,j}$ &  $\rho_{j}$ & $\Delta \widehat{\theta}_{s,j}$ \\ 
  \hline 
  inc     & 0.129 & 0.143 & 0.23   & 3,767 \\ 
  pira    & 0.032 & 0.006 & 0.19   & 449   \\   
  twoearn & 0.015 & 0.011 & -0.07  & -661   \\ 
  \hline
   \hline
\end{tabular}
\end{center}
\small 
\vspace{-.1in}
\caption{Explanatory power of observed covariates in NPM Model. All estimates are debiased and cross-fitted. 
}
\label{table:other x}
\end{table}

\section{Deferred Empirical Example: price elasticity of gasoline demand} 
\label{sec:gasoline}

\subsection{Estimates under conditional ignorability}

An important part of estimating the welfare consequences of price changes is to identify the price elasticity of demand. Here we re-analyze the data on gasoline demand from the 2001 \emph{National Household Travel
Survey} (NHTS) \citep{blundell2012measuring,blundell2017nonparametric,chetverikov2017nonparametric}. This is a household level survey conducted by telephone and complemented by travel diaries and odometer readings (see \cite{blundell2012measuring} for details).  Important variables in the survey include household income, gasoline price, and annual gasoline consumption (as inferred by odometer readings and fuel efficiency of vehicles). Income data corresponds to the median of the income bracket of the household, with $15$ income brackets equally spaced apart in the logarithmic scale. The survey also contains $24$ covariates related to population density, urbanization, demographics and US Census region indicators.\footnote{The data is available on the \textsf{npiv} STATA package \citep{chetverikov2018nonparametric}. The full data contains $3,640$ observations. After applying the same filters suggested by \cite{blundell2017nonparametric} and  \cite{chetverikov2018nonparametric}, the final data contains $3,466$ observations.}

\begin{table}[ht]
\vspace{.1in}
{\centering
\begin{tabular}{lrrcrr}
\hline
\hline
 & \multicolumn{2}{c}{Short Results} & & \multicolumn{2}{c}{Robustness Values}\\
\cmidrule{2-3} \cmidrule{5-6}
Model & Short Estimate & Std. Error & & $\text{RV}_{\theta=-1.5, a=0.05}$ & $\text{RV}_{\theta=0, a=.05}$ \\
\midrule
Partially linear & -0.701 & 0.257 & & 0.026 & 0.019 \\
nonparametric   & -0.761 & 0.360 & & 0.010 & 0.011 \\
\hline
\hline
\end{tabular}}\\
\small
\textbf{Note:} $\rho^2 = 1$; Significance level of 5\%. Standard errors in parenthesis. 
\caption{Minimal sensitivity reporting, gasoline demand.}
\label{tab:gas-robustness-values}
\end{table}

Under the assumption of conditional ignorability, we estimate the average causal derivative of log price on log demand, adjusting for the $24$ observed covariates.\footnote{This can be interpreted as the average price elasticity of demand. We approximate the derivative numerically using a finite difference (e.g., $f'(x) \approx (f(x+0.01) - f(x-0.01))/0.02$).} We consider both a partially linear model, and a fully nonparametric model\footnote{For the partially linear specification we use DML with a cross-validated generic machine learning regression to residualize the outcome and the treatment. For the fully nonparametric specification, we use a generic machine learning approach to estimate both the regression function and the Riesz Representer. In both cases, the regression estimator uses $5-$fold cross-validation to select the best among: (i) lasso models with feature expansions; (ii) random forests; and, (iii) local polynomial forests. The Riesz representer is estimated based on the loss outlined in Remark~\ref{remark:riesz}. We again use $5$-fold cross-validation to choose the best model among a penalized linear Riesz representation with expanded features and a combination of $\ell_1$ and $\ell_2$ penalty \citep{AutoDML,AutoDML-Lasso}, and a random forest representation (ForestRiesz) \citep{riesznet}. In both analyses, in order to reduce the variance that stems from sample splitting for cross-validation and for cross-fitting, we repeat the experiment for $5$ random partitions of the data and average the final estimate, incorporating variation across experiments into the standard error, as described in \cite{DML}. Moreover, since samples are highly correlated within states, we perform grouped cross-validation, where samples of the same state are always in the same fold and we stratify the folds by the census region variable.}. The results are shown in the first column of Table~\ref{tab:gas-robustness-values}. In both models, we obtain estimates similar to the ones obtained in prior literature, with an estimated price elasticity of approximately~$-0.7$.

\subsection{Sensitivity analysis.} Despite having a large number of control variables, there are several reasons why one should worry about the assumption of no unobserved confounders in this setting. For instance, as was argued in \cite{blundell2017nonparametric}, prices vary at the local market level, and unobserved factors that affect consumer preferences could act as unobserved confounders. Another potential source of endogeneity is the fact that we only observe the median of the income bracket of each household, and not the actual income. Since these brackets correspond to large income intervals, the remnant variation in the true income could be another major source of unobserved confounding. This is exacerbated in the larger income brackets, which correspond to larger intervals (and explains the reason why these larger income brackets were not included in prior work).\footnote{Prior work has also analyzed this data via instrumental variable (IV) approaches \citep{blundell2017nonparametric,chetverikov2017nonparametric}, using the distance to the closest major oil platform as an instrument. They find that IV estimates are close to the ones based on unconfoundedness \citep{chetverikov2017nonparametric}. Further, note that the above threats to conditional ignorability are also credible threats to the validity of this proposed instrument. Extensions of our sensitivity results to IV is left to future work.} We thus applied our sensitivity analysis tools to assess the sensitivity of the previous estimates to unobserved confounding.

The second part of Table~\ref{tab:gas-robustness-values} reports the robustness values for price elasticity, such that the sensitivity bounds would contain a target value $\theta$. Here we consider $\theta=-1.5$ (very elastic) and $\theta=0$ (perfectly inelastic). We find that, at the 5\% confidence level,  these robustness values are at around 2\% (PLM) and 1\% (NPM).  These results show that, unless researchers are able to rule out confounding that explains at about  2\% of the residual variation of gasoline price and gasoline consumption, the evidence provided by the data is not strong enough to distinguish between extremes such as a ``very elastic,'' or a ``perfectly inelastic'' demand function. To put this number in context, our coarse measure of income (median of the income bracket) explains around 15\% of the residual variation of gasoline price and 7\% of the residual variation of gasoline demand. It is thus not implausible that remnant variation in the true income could overturn these results.

{
\bibliographystyle{plainnat} 
\bibliography{mybib}
}

\end{document}